\documentclass[twocolumn,showpacs,preprintnumbers,amsmath,amssymb,pra]{revtex4-1}
\usepackage[dvipdfmx]{graphicx}
\usepackage{dcolumn}
\usepackage{bm,color}

\newcommand{\be}{\begin{eqnarray}}
\newcommand{\ee}{\end{eqnarray}}
\newcommand{\nn}{\nonumber\\}
\newcommand{\la}{\langle}
\newcommand{\ra}{\rangle}
\newcommand{\nin}{\noindent}

\def\changed#1{#1}
\def\changedd#1{#1}

\begin{document}


\title{Atomic quantum simulation of a three-dimensional U(1) gauge-Higgs model}

\author{Yoshihito Kuno$^1$, Shinya Sakane$^2$, Kenichi Kasamatsu$^2$, 
Ikuo Ichinose$^1$, and Tetsuo Matsui$^2$}
\affiliation{
$^1$Department of Applied Physics, Nagoya Institute of Technology, 
Nagoya, 466-8555, Japan \\
$^2$Department of Physics, Kindai University, Higashi-Osaka, 577-8502, Japan}

\date{\today}

\begin{abstract}
In this paper, we study theoretically atomic quantum simulations of a U(1) gauge-Higgs model on a three-dimensional (3D) spatial lattice by using an extended Bose-Hubbard model with intersite repulsions on a 3D optical lattice. Here, the phase and density fluctuations of the boson variable on each site of the optical lattice describe the vector potential and the electric field on each link of the gauge-model lattice, respectively. 
The target gauge model is different from the standard Wilson-type U(1) gauge-Higgs model because it has plaquette and Higgs interactions with {\it asymmetric couplings} in the space-time directions. Nevertheless, the corresponding quantum simulation is still important as it provides us with a platform to study unexplored time-dependent phenomena characteristic of each phase in the general gauge-Higgs models.
To determine the phase diagram of the gauge-Higgs model at zero temperature, we perform Monte-Carlo simulations of the corresponding 3+1-dimensional U(1) gauge-Higgs model, and obtain the confinement and Higgs phases.
To investigate the dynamical properties of the gauge-Higgs model, we apply the Gross-Pitaevskii equations to the extended Bose-Hubbard model.
We simulate the time-evolution of an electric flux that initially is put on a straight line connecting two external point charges.
We also calculate the potential energy between this pair of charges and obtain the string tension in the confinement phase.
Finally, we propose a feasible experimental setup for the atomic simulations of this quantum gauge-Higgs model on the 3D optical lattice.
These results may serve as theoretical guides for future experiments.
\end{abstract}

\maketitle

\section{Introduction} \label{intro}

In the last several years, quantum simulation has been one of the most
actively studied subjects in physics \cite{Georgescu}.
Stimulated by the enormous progress made in experimental ultra-cold atomic systems, theoretical proposals have been made for quantum simulations of various physical systems and associated phenomena \cite{book,Blochrev}.
One such proposal is atomic quantum simulation of lattice gauge theories (LGTs) 
\cite{Zohar1,Zohar2,Tagliacozzo1,Banerjee1,Zohar3,Zohar4,Banerjee2,Tagliacozzo2,Zohar5,ours1,Wiese,Zoharrev,Bazavov}.
LGT was introduced by Wilson in 1974 \cite{wilson} to study the mechanism of quark confinement in strong interactions. Since then, various models of LGT have been studied, both analytically and numerically, in various fields of physics, including high energy physics \cite{Rothebook}, condensed matter physics \cite{imrev,strongcor}, neural networks \cite{Takafuji}, etc.
 
The phases studied so far in various models of LGT have been classified as either confinement, Coulomb, or Higgs phases \cite{wilson,Rothebook}.
These phases and the corresponding phase transitions are crucial concepts in various scenes of physical phenomena. 
For LGT models with/without bosonic matter fields, the static equilibrium properties such as the phase diagram can be studied by standard Monte-Carlo (MC) simulations.
On the other hand, for LGTs that include a finite density of fermions, MC simulations generally suffer from the negative-sign problem, and no convincing methods are available to study the static properties.
The atomic quantum simulation does not suffer from the negative-sign problem. Therefore, the realization of quantum simulation has been strongly desired as a way to understand LGT of fermions.
Another, and essential, advantage of quantum simulations of LGT models 
(with either bosons or fermions) is their ability to simulate 
the real-time dynamics (time-development) of the system.
Such simulations can help us not only to study the dynamical properties, 
such as the transport phenomena, but also to intuitively understand the 
characteristics of each of these three phases.
For example, the spatio-temporal images of the electric fluxes provide a visual representation of what happens in each phase.  

The implementation of the local gauge invariance, i.e., the Gauss-law constraint, is a key ingredient in an atomic quantum simulation of LGT.
For the pure compact U(1) LGT, i.e., the theory of self-interacting compact U(1) gauge fields without matter fields, the Gauss-law constraint is expressed by the operator identity $\sum_{i=1}^3\nabla_i \hat{E}_{r,i}=0$, where $\nabla_i$ is the lattice difference operator 
\changed{($\nabla_i f_r\equiv f_{r+\hat{i}}-f_r$)} 
and $\hat{E}_{r,i}$ is the electric field operator on the link [See Eq.~(\ref{dive}) below].
Some proposals \cite{Zohar1, Zohar2} have appeared to implement this Gauss law in cold atom systems. 
\changed{However, they must necessarily take a particular limit  
expressed as ``$\gamma \to 0$" for the strength $\gamma^{-2}$ of certain set of interactions between atoms (See Table I for the definition of $\gamma$)} \cite{gamma}. 
This limit seems hard to achieve experimentally, but the available atomic systems without this limit 
\changed{give rise to $\sum_i\nabla_i \hat{E}_{r,i}\propto \gamma (\neq 0)$} and 
certainly break the local gauge symmetry.

\changed{
In the previous work \cite{ours1}, we 
started with the so called extended Bose-Hubbard model, which 
is given by adding off-site interactions to the Bose-Hubbard model.
 In the path  from the extended Bose-Hubbard
model to a would-be gauge theory, we encountered 
a $\gamma$-dependent term in the Hamiltonian [see Eq.~(\ref{HEBH2})],
which explicitly breaks the gauge symmetry of the pure gauge theory
as in the models discussed in Refs.~\cite{Zohar1,Zohar2}.
In stead of taking the limit $\gamma\to 0$ to obtain the pure gauge theory,  
we introduced a complex scalar field $\phi(x)$ (Higgs field) and regarded this term as 
an interaction term between the gauge field
and the Higgs field. The relation  
$\sum_i\nabla_i \hat{E}_{r,i}\propto \gamma$ now represents the genuine 
Gauss law where the right-hand side  is nothing but 
the charge of this Higgs field. 
For this purpose, 
$\phi(x)$ should appear in its trivial form $\phi(x)=1$, that is, (i)  in the 
London limit, i.e.,
its radial fluctuations are frozen as $\phi(x) = \exp(i \varphi(x))$ ($|\phi(x)|=1$),  {\it and} 
(ii)  in the particular gauge $\varphi(x) = 0$ ($\phi(x)=|\phi(x)|$) which is called the unitary gauge.  
}

\changed{
Fixing a gauge is a justified procedure because we are interested in a U(1) 
gauge theory,
and  the expectation value of any gauge-invariant quantity 
and the related quantities such as the phase diagram are independent of the gauge 
that one fixes \cite{wilson}.
In this way, one may study
a U(1) gauge theory with Higgs field (in the London limit) for general $\gamma$.
This argument is general and applies to models similar to the extended Bose-Hubbard model in arbitrary dimensions.
}

Next, we comment on our treatment of the Higgs field in the London limit (freezing radial fluctuations). 
As is well known, the unified theories of elementary particles, such as the
Weinberg-Salam theory, are gauge theories containing elementary fermions and Higgs bosons.
Since Wilson's introduction of LGT \cite{wilson} for the strong interaction, many gauge-Higgs models have also been formulated as LGT and studied intensively, both with and without radial degrees of freedom $|\phi_x|$ \cite{FS}. 
These studies have proved that the models in the London limit are widely accepted as interesting models, because they describe the low-energy phase dynamics faithfully (note that the radial excitations are massive) and exhibit interesting phase structures and gauge-field properties such as
the Anderson-Higgs mechanism.
In particular, the study of the Higgs-confinement phase transition by these LGT models is important because such a phase transition is expected to have taken place in the early universe.

We list the relations between the original atomic model and the target gauge model studied in Ref.~\cite{ours1}.
The atomic model is the extended Bose-Hubbard model on a 3D optical lattice and the effective gauge model at low energies is the gauge-Higgs model on a 3D lattice (we call it the gauge lattice).

\nin
(A1) The site $a$ of the 3D optical lattice is the midpoint of the link $(r,r+\hat{i})$ of the gauge lattice, where $r$ is the site of the gauge lattice and $\hat{i}\ (i=1,2,3)$ is the unit lattice vector in the positive $i$-th direction (See Fig.~\ref{bct_lattice}).\\
\nin
(A2) The phase $\hat{\theta}_a$ of the bosonic operator $\hat{\psi}_a=\exp(i\hat{\theta}_a) \sqrt{\hat{\rho}_a}$ for the bosons sitting on site $a$ is identified with the U(1) gauge field 
$\hat{\theta}_{r,i}$ on the link $(r,r+\hat{i})$. 
This guarantees the U(1) periodicity of the gauge-Higgs model (compactness) under $\hat{\theta}_{r,i}\to \hat{\theta}_{r,i}+2\pi$.
In LGT, the electric field operator $\hat{E}_{r,i}$ is conjugate to the vector potential $\hat{\theta}_{r,i}$. 
The above identification implies that $\hat{E}_{r,i}$ corresponds to the amplitude operator $\sqrt{\hat{\rho}_a}$ of atoms.\\
\nin
(A3) We assumed that, on an average, the density of atoms $\hat{\rho}_a$ has a uniform distribution, that $\la\hat{\rho}_a\ra=\rho_0$, and that its fluctuation $\hat{\eta}_a\equiv  \hat{\rho}_a-\rho_0$ is small compared to $\rho_0$, $\hat{\eta}_a/\rho_0 \ll 1 $ at low energies.
So, we neglect higher-order terms than $O((\hat{\eta}_a/\rho_0)^2)$ in the effective action. 
In practice, these conditions may suggest $\rho_0 \gtrsim 10$, which is achieved in a relatively easy manner in experiments (See Sec.~\ref{formulation} for details). 
We stress that we {\it do not} assume the Bose-Einstein condensation (BEC)
of cold atoms a priori; rather we are interested in the transition itself from a disordered incoherent state to a BEC, because a BEC transition corresponds to a confinement-deconfinement transition of a gauge theory (See Sec.~\ref{phasedia}). 

Then the explicit relationship between the two models is established; the interaction parameters of the gauge-Higgs model are given by explicit functions of the parameters of the extended Bose-Hubbard model. 
In Ref.~\cite{ours1} we obtained the phase diagram of the corresponding 
(3+1)D U(1) gauge-Higgs model defined on a (3+1)D lattice \cite{mc3+1d}
for some set of parameters. 
This phase diagram may be used as a guide for experimentalists to select 
parameters for the extended Bose-Hubbard model, i.e., parameters for experimental setups of quantum simulations.
Recently, the extended Bose-Hubbard model has been realized for cold atoms on a 3D optical lattice and some interesting experimental results have been reported \cite{expebhm}.
We are looking forward to hearing about further results of experimental studies, especially those that are relevant to LGTs that include the gauge-Higgs model studied in the present paper. 

Let us point out that our way to introduce the U(1) gauge field by the points (A2) and (A3) is in strong contrast to another way 
\cite{Zohar2,Tagliacozzo1,Banerjee1,Zohar3,Zohar4,Banerjee2,Tagliacozzo2,Zohar5} using the quantum link model (gauge magnet).
The gauge-magnet recipe for U(1) gauge operator prepares a multiplet of boson states $|S_{az}\ra$ at each site $a$ with the multiplicity $2S+1$, and uses the pseudo-spin formalism.
Then the electric field $\hat{E}_{a}$ is identified as $\hat{E}_{a}=\hat{S}_{az}$. Therefore, the eigenvalue $E_{a} = S_{az}$ is restricted to the range $(-S,S)$.
To recover the expected genuine support of the U(1) momentum operator $E_a \in \bf{Z}$ (integer) one needs to take the limit $S\to \infty$.  
We note that, in Ref.~\cite{Zohar5}, it is proposed that the gauge symmetry is implemented by angular-momentum conservation in the scattering processes between a matter particle and a boson. 
However, the gauge field is a composite of two bosons and is {\it not} a genuine U(1) field $\exp(i\theta)$ for finite $S$. 

In our second paper \cite{ours2}, we focused on the extended Bose-Hubbard model in the two-dimensional (2D) optical lattice and the resulting 2D gauge-Higgs model.
A reason for choosing the 2D system is that it is easier to set up 
experimentally than the 3D system. 
We studied the following three points: 

\nin (B1)
Phase diagram of the (2+1)D gauge-Higgs model as well as the extended Bose-Hubbard model itself; we found that the Coulomb phase is missing, as expected, from the study of the related models \cite{polyakov}.  

\nin (B2)
Formulation and solution of the Gross-Pitaevskii equation (GPE) \cite{latticegpe,gpe,gpe2} of the extended Bose-Hubbard model; GPE \cite{gpe0} is an approximate but useful equation describing the time-evolution of a quantum system. 
It has been applied widely, mainly in condensed matter physics \cite{gpe,gpe2}.

\nin
(B3)
Proposal of two feasible methods to set up a practical atomic simulator;
one is based upon the excited bands of an optical lattice and the other uses dipolar atoms in a triple-layer optical lattice.
These may help experimentalists to set up their systems for a quantum simulation of LGT.

In this paper, we return to the 3D gauge-Higgs model again, and 
present a detailed account of the first paper \cite{ours1}.
Furthermore, we study the following three new aspects:\\
\nin
(C1) We refine and generalize the phase structure.\\
\nin
(C2) We extend the GPE study of dynamical properties made for the 2D gauge-Higgs model.\\
\nin
(C3) We propose a feasible experimental set up of a system describing the 3D gauge-Higgs model. 

The structure of the paper is as follows.
In Sec.~\ref{formulation}, we introduce the 3D extended Bose-Hubbard model, the Bose-Hubbard model with intersite interactions, and explain how the gauge-Higgs model appears as its effective model at low energies (i.e., at low temperatures).
In Sec.~\ref{phasedia}, we discuss the results of the MC simulations of the resultant gauge-Higgs model.
The phase diagrams are shown and the physical properties of each phase are explained.
The Higgs phase in the gauge theory corresponds to the superfluid phase of cold atoms and the confinement phase corresponds to the Mott-insulator phase, which has no phase coherence. 

In Sec.~\ref{gpdynamics}, we study the dynamical properties of the gauge-Higgs model by using GPE.
In particular, we are interested in the time evolution of an electric flux
put on the links of the gauge lattice.
The electric flux behaves quite differently in the confinement and Higgs phases.
The string tension of the electric flux is also calculated.
In Sec.~\ref{expproposal}, we propose feasible experiments of the extended Bose-Hubbard model to simulate the gauge-Higgs model. 
The recipe starts from a system of two species of bosons (A and B atoms) and subsequently changes over the optical-lattice structure to obtain the desired system consisting only of the A atoms. 
Section~\ref{concle} is devoted to the conclusions.

Our paper describes a somewhat lengthy and subtle analysis, but this is necessary to make it self-contained. 
In Sec.~\ref{formulation}, the important point is that the Hamiltonian of the suitably designed atomic system Eq.~(\ref{HEBH2}) or Eq.~(\ref{HGHM}) (after the transformation Eq.~(\ref{AE})) can be identified by the gauge-Higgs model Eq.~(\ref{HGHM}), whose partition function is given by Eq.~(\ref{4DGHM}). 
Readers who are not interested in the details of the statistical analysis of the Monte-Carlo simulations in Sec.~\ref{phasedia}, can skip this section, except for the phase diagrams of Figs.~\ref{PD} and \ref{PD2}. 
The contents of Secs.~\ref{gpdynamics} and~\ref{expproposal} would be the parts that, we expect, attract most of the readers, because these parts are closely related to the experimental observations. 

To close this section, let us confirm the motivation for a quantum simulation of this gauge-Higgs model.
First, it is out of criticism that the Wilson-type models \cite{wilson} of LGT (which are well known and studied mainly in high-energy physics), are the primary targets of quantum simulations, because their time dependent behavior is certainly of great interest (For a partial list of explicit fields of application, see Ref.~\cite{ours1}).
These models have symmetric couplings in the space-time directions, thus reflecting relativistic invariance. In contrast, the present gauge-Higgs model has asymmetric plaquette and Higgs couplings [see Eq.~(\ref{cs}) below], reflecting that the starting point, the extended Bose-Hubbard model is nonrelativistic. However, quantum simulation of the gauge-Higgs model itself does not lose its importance. 
As one reason for its existence and usefulness, one may first list that 
there are currently no realistic proposals available to simulate the Wilson-type U(1) gauge-Higgs model, i.e., with symmetric plaquette couplings with arbitrary strength, and with symmetric Higgs couplings.  
This may sound like a passive reason, but we recall that we have very little solid knowledge of time-dependent quantum phenomena of 3D gauge theory. 
The present gauge-Higgs model exhibits both the confinement and Higgs phases as we will see in Sec.~III. 
Any experimental information of the time-dependent phenomena of this model, such as the motion of electric flux in each phase, as we consider in Sec.~IV, is thus welcome. 
Second, we list and stress the importance of quantum simulation 
of nonrelativistic models of LGT themselves in condensed matter physics. 
In this field, the gauge-theoretical approach has proved to be a powerful method to understand and describe physical phenomena \cite{imrev}, 
especially for systems with strong correlations \cite{strongcor}. 
The explicit results for the time development of an electric flux obtained in Sec.~IV and in Ref.~\cite{ours2} seem to support well these reasons for the importance of quantum simulation of nonrelativistic gauge models.

\section{From the extended Bose-Hubbard model to the U(1) gauge-Higgs model} \label{formulation}

\begin{figure}[t]
\centering
\includegraphics[width=7.0cm]{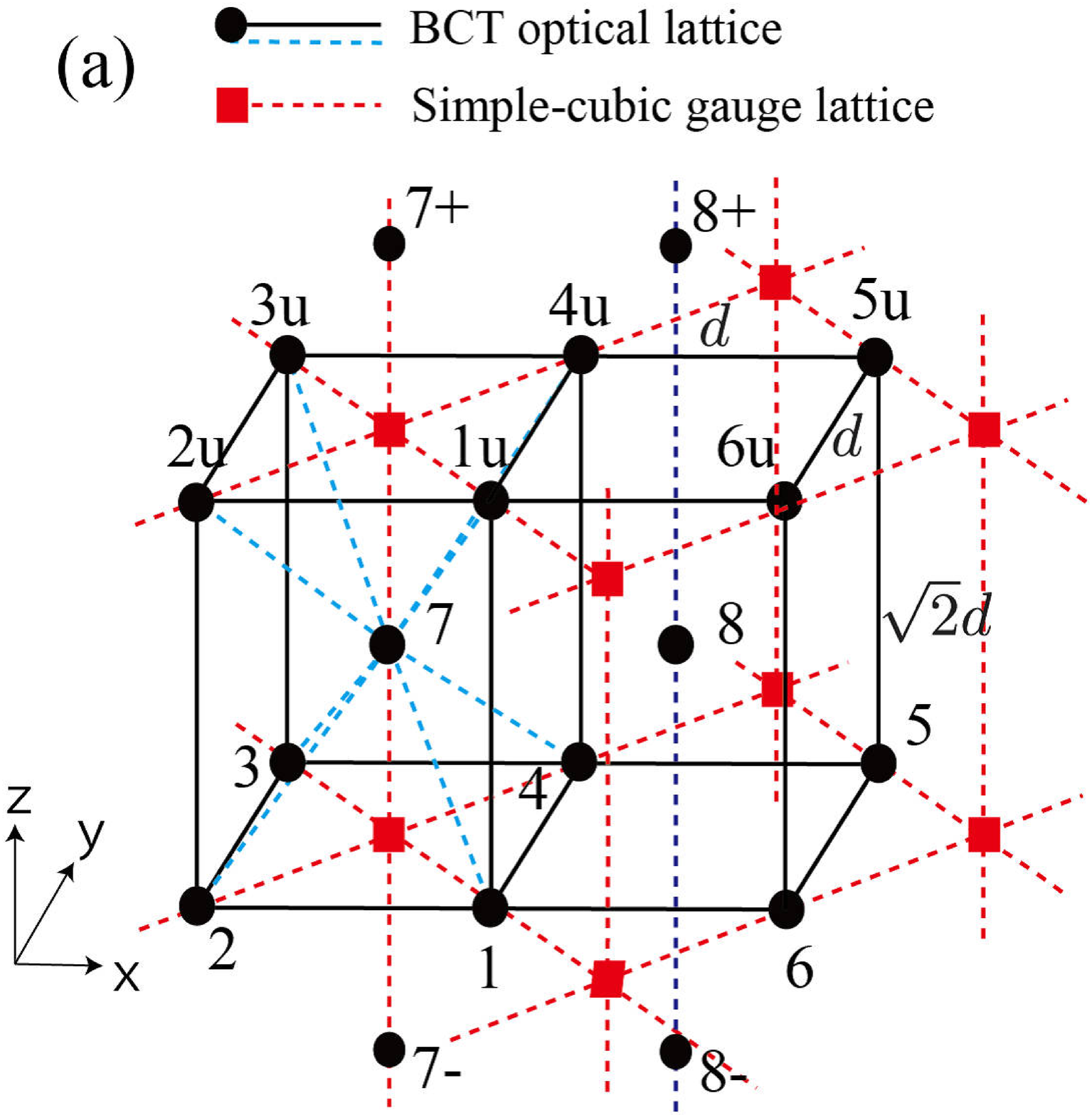}\\
\hspace{-.8cm}\includegraphics[width=6.0cm]{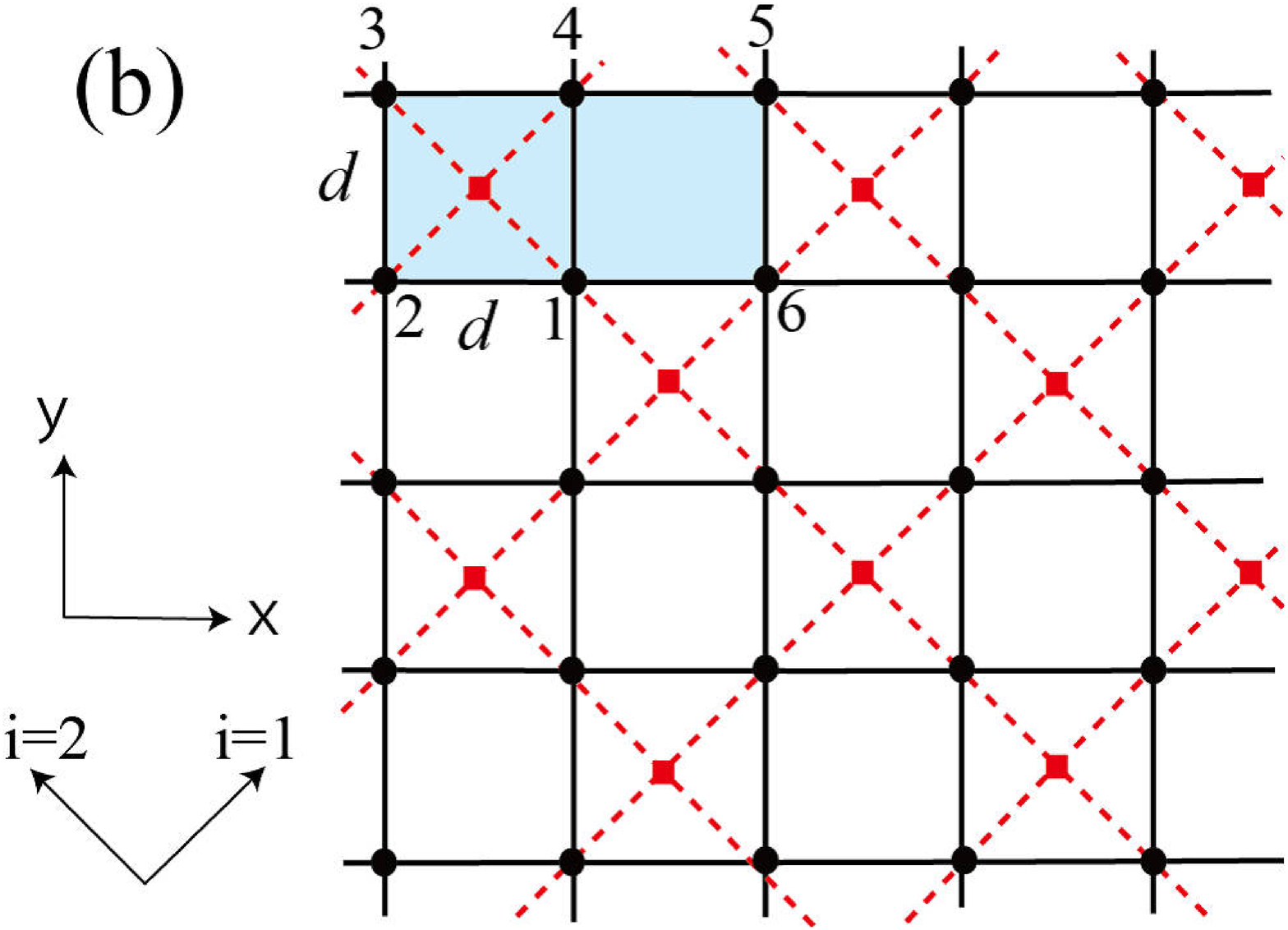}
\caption{(Color online) Sites of the BCT optical lattice (black circles) and the simple-cubic gauge lattice (red squares).  
(a) Cuboid having the eight vertices $1,2,3,4,1_u,2_u,3_u,4_u$, the center 7, and dimensions $d\times d\times \sqrt{2}d$. 
This is a unit cell of the BCT optical lattice. 
These sites are potential minima and cold atoms reside on them.
The cube having eight vertices (red squares) and dimensions $(\sqrt{2}d)^3$ is a unit cell of the simple-cubic gauge lattice on which the gauge-Higgs model is defined.
(b) The cross section in the $x$-$y$ plane ($i=1,2$ plane in the gauge lattice). 
The combined 3D system of the optical lattice and the gauge lattice is obtained by piling up copies of this cross section along the $z$ ($i=3$) direction with equal distance $\sqrt{2}d$ and further inserting between them copies of another kind of cross section, which is made of only the centers of the cuboids of the BCT lattice [such as 7 and 8 in (a)].
The essence of the relation between these two lattices is that each black circle sits on the midpoint of a nearest-neighbor pair of red squares, i.e., a site of the optical lattice corresponds to a link of the gauge lattice. 
}
\label{bct_lattice}
\end{figure}

\begin{table*}[t]
\centering
\caption{\label{paratable} Atomic parameters $J_{ab}$ and $V_{ab}$ in Eq.~(\ref{HEBH}) for the BCT optical lattice of Fig.~\ref{bct_lattice}. 
The parameters for the pairs $(a,b)$ that have longer distance than NNN 
are set zero. 
The pairs that belong to each of the (sub)groups (i), (ii), and (iv) are 
illustrated in Fig.~\ref{table1-2}.
With this choice of parameters, the extended Bose-Hubbard model can be equivalent to the 3D gauge-Higgs model.
In particular, group (iv) is responsible for Gauss law [See Eq.~(\ref{dive}) below].} 

\renewcommand{\arraystretch}{1.25}
\begin{tabular}{p{1cm}p{4cm}cp{8.5cm}ccc}\hline
group&pairs in each group&&$(a,b)$&& $J_{ab}$& $V_{ab}$\\ \hline
(i)&NN1 and NN2 in an even unit cell (center site 7)&&(1,2), (2,3), (3,4), (4,1), (1u,2u), 
(2u,3u), (3u,4u), (4u,1u), 
(1,7), (2,7), (3,7), (4,7), (1u,7), (2u,7), (3u,7), (4u,7)&& $J$&  
$\gamma^{-2}$\\ \hline
(ii)&NN1 in an odd unit cell (center site 8)&&(1,6), (6,5), (5,4), (4,1), 
(1u,6u), (6u,5u), (5u,4u), (4u,1u)&& $J$&  
$\gamma^{-2}$\\ \hline
(iii)&NN2 in an odd unit cell (center site 8)&&(1,8), (4,8), (5,8), (6,8), (1u,8), (4u,8), 
(5u,8), (6u,8)&& 0&0\\ \hline
(iv)& NNN1 with a mid-point gauge lattice site&&(1,3), (2,4), (1u,3u), (2u,4u), (7,7+), (7,7-)&& 0&  
$\gamma^{-2}$\\ \hline 
(v)& NNN2 with no mid-point gauge lattice site&&(1,5), (4,6), (1u,5u), (4u,6u), (1,1u), (2,2u), (3,3u), (4,4u), (5,5u), (6,6u), (8,8+), (8,8-)&& 0 &0 \\ \hline
\label{vab}
\end{tabular}\\
\end{table*}
\begin{figure}[ht]
\centering
\includegraphics[width=6cm]{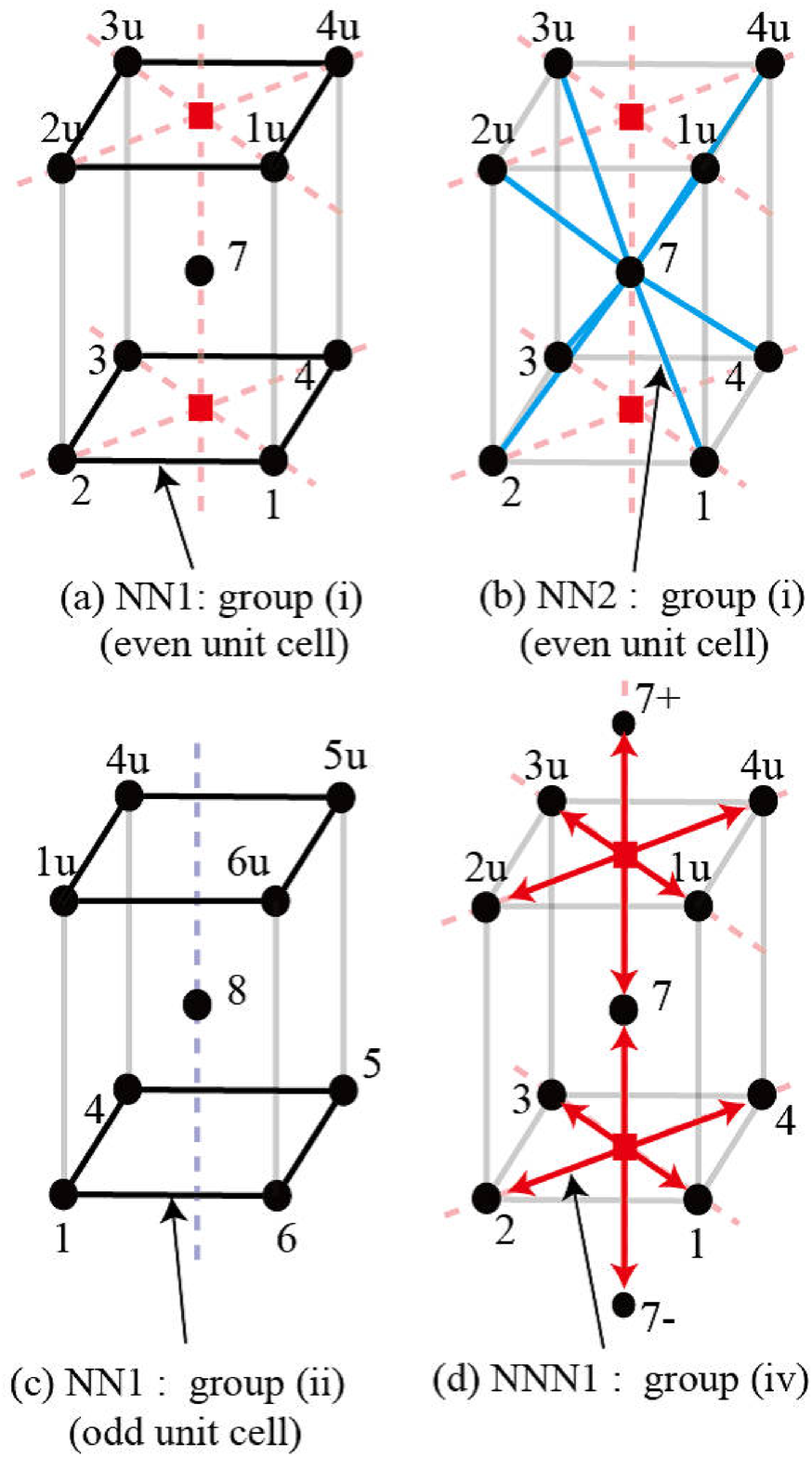}
\caption{(Color online) 
Illustration of pairs of sites in each (sub)group in Table \ref{paratable} with nonvanishing $J_{ab}$ and/or $V_{ab}$.
(a) NN1 of group (i), (b) NN2 in group (i), (c) NN1 in group (ii), and (d) NNN1 in group (iv).  
}
\label{table1-2}
\end{figure}

In this section, we start from the extended Bose-Hubbard model that is to be realized by ultra-cold atom systems on an optical lattice and show that it is equivalent to the 3D gauge-Higgs model under certain conditions.  
For the optical lattice, we choose the body-centered tetragonal (BCT) lattice where the unit cell is a cuboid of size $d_x\times d_y\times d_z$ with $d_x=d_y=d$ and $d_z=\sqrt{2}d$.
This BCT lattice is illustrated in Fig.~\ref{bct_lattice}, where the (black) circles denote its sites. 
These sites are the potential minima and atoms may sit on them (See Sec.~\ref{expproposal} for details).
In Fig.~\ref{bct_lattice}, we also draw the gauge lattice on which the gauge-Higgs model is defined; its sites are shown by (red) squares.
The gauge lattice is a simple cubic lattice with the lattice spacing $\sqrt{2}d$, so the volume of unit cell $(\sqrt{2}d)^3$ is twice as large as that of $d^2\times \sqrt{2}d$. 
Every optical lattice site sits on the midpoint of a nearest-neighbor (NN) pair of sites of the gauge lattice, i.e., it sits on a link of the gauge lattice.
As long as one imposes that (i) the 3D gauge lattice is simple cubic and (ii) a link of the gauge lattice corresponds to a site of the optical lattice, the optical lattice has to be BCT as shown in Fig.~\ref{bct_lattice}.

The Hamiltonian of the extended Bose Hubbard model on this BCT lattice is given by
\begin{eqnarray}
H_{\rm EBH}&=&-\sum_{a \neq b}J_{ab}\hat{\psi}^\dagger_a\hat{\psi}_b
+{V_0\over 4}\sum_a\hat{\rho}_a(\hat{\rho}_a-1) \nonumber \\
&&+\sum_{a \neq b}{V_{ab} \over 2}\hat{\rho}_a\hat{\rho}_b,
\label{HEBH}
\end{eqnarray}
where $a$ and $b$ denote the sites of the BCT lattice,
$\hat{\psi}_a \ (\hat{\psi}^\dagger_a)$ is the annihilation (creation) operator of the bosonic atom at site $a$ satisfying the canonical commutation relation $[\hat{\psi}_a, \hat{\psi}^\dagger_b]=\delta_{ab}$, 
and $\hat{\rho}_a=\hat{\psi}^\dagger_a\hat{\psi}_a$ is the atomic density. 
The parameter $J_{ab} (=J_{ba})$ is the hopping amplitude between the pair
of sites $a$ and $b$, 
whereas $V_0$ and $V_{ab} (=V_{ba})$ are the on-site and off-site interactions between atoms at $a$ and $(a,b)$, respectively.
In this work, we confine ourselves to the repulsive interactions $V_0>0$ and $V_{ab}>0$.

The values of $J_{ab}$ and $V_{ab}$ are shown in Table \ref{vab}.
To explain it, it is useful to partition unit cells of the optical lattice into ``even"-column cells (as the ones with the center sites 7, 7$_{\pm}$ in Fig.~\ref{bct_lattice}) and ``odd"-column cells (as the ones with 8, 8$_{\pm})$, where each column of cells extends in the $z$-direction. 
Even and odd columns are distinguished by their signature $(-)^{x+y}$ and face each other alternatively as in a black-red checker board in the $x$-$y$ plane.

The NN pairs in the BCT lattice of Fig.~\ref{bct_lattice} have a distance $d$ and consist of two types: NN1 --- the ordinary pairs connecting each corner in the $x$-$y$ plane [as (1,2) in Fig.~\ref{table1-2} (a) and (1,6) in Fig.~\ref{table1-2} (c)] and NN2 --- the pairs starting from each body-center site [as (1,7) in Fig.~\ref{table1-2} (b)] because of the choice $d_3=\sqrt{2}d$. 
Then, $J_{ab}$ and $V_{ab}$ have the non-vanishing values $J_{ab}=J$ and
$V_{ab}=\gamma^{-2}$ for NN1 in all the unit cells and for NN2 in all the even cells; and $J_{ab}=0$, $V_{ab}=0$ for NN2 in all the odd cells.
$J_{ab}$ are truncated up to the NN pairs and $J_{ab}=0$ for longer pairs.

The next-NN (NNN) pairs have the distance $\sqrt{2}d$, and are classified in two types: 
NNN1 --- the pair has a gauge lattice site at its midpoint (such as (2,4) in Fig.~\ref{table1-2} (d)) and
NNN2 --- the pair does not have a gauge lattice site at its midpoint (such as (1,5) in Fig.~\ref{bct_lattice}).
Then, $V_{ab}$ is non-vanishing as $V_{ab}=\gamma^{-2}$ for NNN1, while $V_{ab}=0$ for NNN2. 
Here, $V_{ab}$ are truncated up to the NNN pairs and $V_{ab}=0$ for longer pairs.

These settings of parameters might seem rather strange, e.g., some NN and NNN pairs have the same value $V_{ab}=\gamma^{-2}$ \cite{nnvsnnn}. 
However, it is necessary to relate this extended Bose-Hubbard model to a model of LGT, and in Sec.~\ref{expproposal} we present a feasible experimental way to set up an atomic system that describes the extended Bose-Hubbard model with this choice of parameters.

To derive the effective gauge field theory from $H_{\rm EBH}$ in
Eq.~(\ref{HEBH}), we introduce an operator corresponding to the phase 
degrees of freedom $\hat{\theta}_a$ of $\hat{\psi}_a$ as
\begin{equation}
\hat{\psi}_a=e^{i\hat{\theta}_a}\sqrt{\hat{\rho}_a}.
\label{phase}
\end{equation}
Then $\hat{\rho}_a$ and $\hat{\theta}_a$ are conjugate to each other, 
satisfying the canonical commutation relation,
$[\hat{\rho}_a, \hat{\theta}_b]=i\delta_{ab}$.
We furthermore separate $\hat{\rho}_a$ into its mean value $\bar{\rho}_a$
and the quantum fluctuation $\hat{\eta}_a$ as 
\be
\hat{\rho}_a=\bar{\rho}_a+\hat{\eta}_a.
\ee
The value $\bar{\rho}_a$ may be estimated in various ways: some mean-field theory, more elaborated methods making use of self-consistency,
and/or numerical simulations. 
In this paper, we consider the case in which a homogeneous state $\bar{\rho}_{a}=\rho_0$ ($a$-independent value) is realized.
In Appendix \ref{mftdenrho0}, we study a simple mean-field theory to determine $\bar{\rho}_{a}$ for the choice of Table \ref{vab}.
We show there that the inhomogeneous states compete in energy with the homogeneous state, but for sufficiently large on-site repulsion $V_0$ compared to the inter-site repulsion $\gamma^{-2}$, the homogeneous state has lower energy as expected. 
As described in Appendix \ref{mftdenrho0}, the homogeneous state is stable for $\gamma^{-2}/V_0 \alt 0.5$ for sufficiently large average atomic density per site.
For the parameters that support the inhomogeneous state, one may still have a chance to obtain a lattice gauge model from the extended Bose-Hubbard model of Eq.~(\ref{HEBH}).
For this purpose, however, the parameters $J_{ab}$ and $V_{ab}$ should be altered from the values in Table \ref{vab} to those that reflect no uniformity of $\bar{\rho}_a$. 
In short, Gauss law relates $J_{ab}$ not only to $V_{ab}$ but also to $\bar{\rho}_a$.  

We identify each site $a$ of the original BCT optical lattice as a link $(r,i)\equiv (r,r+i)$ of the cubic gauge lattice on which the gauge model is defined. 
Here, $r =(x_1,x_2,x_3)$ is the site of the gauge lattice and $i =1,2,3$ is the direction index (Below we also use $i$ as the unit vector $\hat{i}$). 
We take the directions $i=1,2$ as shown in Fig.~\ref{bct_lattice} (b) and $i=3$ to the $z$-direction.

By setting $\bar{\rho}_a=\rho_0$, choosing the parameters in the Hamiltonian $H_{\rm EBH}$ in Eq.~(\ref{HEBH}) according to Table \ref{vab},
and expanding the density operator in powers of $\hat{\eta}_a$ up to the second order, $H_{\rm EBH}$ of Eq.~(\ref{HEBH}) becomes
\be
&&H_{\rm EBH}|_{\rm Table\ I}=H'_{\rm EBH} +O((\hat{\eta}/\rho_0)^3),\nn
&&H'_{\rm EBH}=\sum_r\left[{1 \over 2\gamma^2}\Big(
\sum_k\hat{\eta}_k\Big)^2 
+{V'_0 \over 2}\sum_k(\hat{\eta}_k)^2\right.  \nonumber \\
&&\hspace{1.2cm}\left.-{\rho}_{0}J\sum_{(m,n)}
\cos(\hat{\theta}_m-\hat{\theta}_n)\right],\ V_0'\equiv V_0-\frac{2}{\gamma^2},
\label{HEBH2}
\ee
where $k=1,2,3,4,7,7_{\_}$ represents the six optical lattice sites surrounding $r$ with the distance $d/\sqrt{2}$, and $(m,n)$ 
= (1,2), (2,3), (3,4), (4,1), (7,1), (7,2), (7,3), (7,4),
(7$_\_$,1), (7$_\_$,2), (7$_\_$,3), (7$_\_$,4),
are the twelve NN optical lattice pairs surrounding $r$.
We note that the condition of homogeneous state $\gamma^{-2}/V_0 \lesssim 0.5 $ implies that $V_0' > 0$.

Let us make some comments on $H'_{\rm EBH}$. 
As explained in Appendix \ref{mftdenrho0}, the average value $\rho_0$ is adjusted so that no linear terms in $\hat{\eta}_a$ appear in $H'_{\rm EBH}$. 
Further, a straightforward expansion of $H_{\rm EBH}$ up to $O(\hat{\eta}^2)$ gives rise to an extra $J$-term $\propto J \hat{\eta}^\dag_m\hat{\eta}_n\exp[i(\theta_m-\theta_n)]+{\rm H.c.}$.
We neglected this term because our main interest is the BEC transition of cold atoms, since this transition corresponds to the confinement-deconfinement transition in gauge theory (see Sec.~\ref{phasedia}).
Due to the limited accessibility to extremely low temperatures, the average density $\rho_0$, which generally increases as the transition temperature rises, cannot be set arbitrary small. 
One expects, e.g., that $\rho_0 \gtrsim 10$ in practical experiments.
Then the extra $J$ term can be neglected due to an extra suppression factor $\rho_0^{-1} (\lesssim 10^{-1})$ compared to the last $\rho J$-term in $H'_{\rm EBH}$.
We note that the coefficients $\gamma^{-2}$ and $V_0'$ of the remaining two terms in $H'_{\rm EBH}$ should compete with $\rho_0 J$ near the phase transition, i.e., these three parameters are roughly of the same order.
This point is confirmed a posteriori in the phase diagram Fig.~\ref{PD2} in Sec.~\ref{phasedia}.

In the U(1) gauge theory, the vector potential $\hat{\theta}_{r,i}$ and the electric field $\hat{E}_{r,i}$ on the link $(r,i)$ are a set of canonically conjugate operators satisfying $[\hat{E}_{r,i}, \hat{\theta}_{r',i'}]=-i
\delta_{rr'}\delta_{ii'}$ \cite{ks}.
They have the eigenvalues $\theta_{r,i} \in [0, 2\pi)\ {\rm mod} (2\pi)$ and $E_{r,i} =0,\pm1,\pm2,\cdots \in\ {\rm Z}$ as mentioned in Sec.~\ref{intro}.
Therefore, we identify  
\begin{equation}
\hat{\theta}_{r,i}\equiv (-)^r\hat{\theta}_a, \;\; 
\hat{E}_{r,i} \equiv -(-)^r\hat{\eta}_a,
\label{AE}
\end{equation}
where $(-)^r=(-)^{x_1+x_2+x_3}$ (we use the same letter $\theta$ for the optical and gauge lattice).   
It is straightforward to check that $\hat{\theta}_{r,i}$ and $\hat{E}_{r,i}$ are a canonical pair.
The sign factor $(-)^r$ plays a crucial role to obtain the Gauss-law
equation as seen below. 
At this stage, we stress that the behavior of the electric field such as motion of the electric fluxes is simulated by observing the density 
fluctuations of the extended Bose-Hubbard model experimentally. 

\begin{figure}[t]
\centering
\includegraphics[width=8cm]{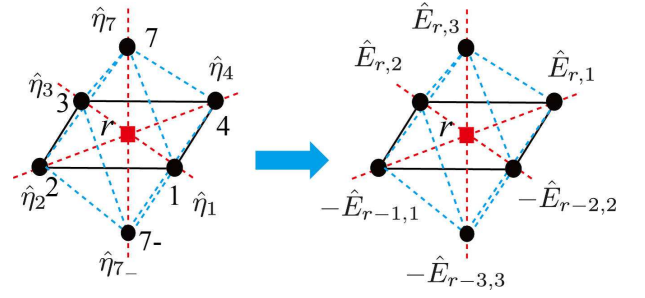}
\caption{(Color online) 
Illustration of $\sum_k \hat{\eta}_k$ in Eq.~(\ref{HEBH2}). 
$k$ is the six optical-lattice sites NN corresponding to a gauge-lattice site $r$. 
The relation between the density fluctuation $\hat{\eta}_k$ and the electric field of Eq.~(\ref{AE}) (The figure is for the odd site $r$ with $(-)^r=-1$) converts the summation $\sum_k \hat{\eta}_k$ of the lattice divergence of electric fields $\hat{E}_{r,i}$ into $\sum_k \hat{\eta}_k=\sum_i(\hat{E}_{r,i}-\hat{E}_{r-i,i})=\sum_i{\nabla_i}
\hat{E}_{r,i}\propto $ div$\vec{E}(\vec{r})$. 
}
\label{sum_k}
\end{figure}

By rewriting $H'_{\rm EBH}$ in Eq.~(\ref{HEBH2}) in terms of $\hat{\theta}_{r,i}$ and $\hat{E}_{r,i}$, and sorting some terms suitably,
we obtain the following Hamiltonian $H_{\rm GH}$ of the gauge-Higgs model: 
\begin{eqnarray}
\hspace{-0.3cm}
H_{\rm GH}&\equiv& H'_{\rm EBH}=\sum_r H_r\nn
H_r&=&{1 \over 2\gamma^2}\Big[
\sum_i(\hat{E}_{r,i}-\hat{E}_{r-i,i})\Big]^2 
+{V'_0 \over 2}\sum_{i}\hat{E}^2_{r,i}  \nonumber \\
&&-{\rho}_{0}J\sum_{i<j}
\left[\cos(\hat{\theta}_{r,i}-\hat{\theta}_{r,j})+
\cos(\hat{\theta}_{r,i}+\hat{\theta}_{r+i,j})\right.\nn
&&\left.+\cos(\hat{\theta}_{r+i,j}-\hat{\theta}_{r+j,i})
+\cos(\hat{\theta}_{r,j}+\hat{\theta}_{r+j,i})
\right].
\label{HGHM}
\end{eqnarray}
The first $\gamma^{-2}$-term in $H_{\rm GH}$ is just a rewriting of the first term of  $H'_{\rm EBH}$ of Eq.~(\ref{HEBH2}) by using Eq.~(\ref{AE}) as $\sum_k\hat{\eta}_k 
 =\sum_i(\hat{E}_{r,i}-\hat{E}_{r-i,i})$.
This is illustrated in Fig.~\ref{sum_k}. 
The last $\rho_0 J$-term in $H_{\rm GH}$ comes from the NN hopping term of the extended Bose-Hubbard model and represents the interaction between two phases put on the two links on the gauge lattice such as $(r,i)$ and $(r,j)$.
These two links have a common gauge lattice site ($r$) and make a right angle, forming an L-shape.
The four terms in the square bracket represent the four L-shapes, and they are four pieces contained in the plaquette $(r,r+i,r+i+j,r+j)$ of the gauge lattice. 
The relative signature between the two phases is determined by the factor $(-)^r$ in Eq.~(\ref{AE}).

To reveal that the system (\ref{HGHM}) can be regarded as a gauge system, let us investigate Eq.~(\ref{HGHM}) term by term.
The first term describes Gauss law.
This term can be regarded as the Gaussian distribution of the divergence of $E_{r,i}$.
In fact, its coefficient $(2\gamma^2)^{-1}$ determines the variance as $\gamma^{2}$.
Therefore, the expectation value $E_{r,i}$ of $\hat{E}_{r,i}$ can be estimated as
\be
\left|\sum_i\nabla_i E_{r,i}\right| \lesssim \gamma,\quad \nabla_i E_{r,i} \equiv E_{r,i}-E_{r-i.i}.
\label{dive}
\ee
We note that $\sum_i \nabla_i E_{r,i}$ has the continuum limit
$\propto {\rm div}\vec{E}(r)$ as $d\to 0$ \cite{ks}. 
Equation (\ref{dive}) is Gauss law on the gauge lattice with a matter-field charge density $\propto \pm\gamma$. 
In the path-integral formulation below (using real time instead of imaginary time), we will identify this matter field with a complex scalar field $\hat{\phi}_r$ (charged Higgs field) in the London limit. 
By taking the limit $\gamma^2\to 0$, Eq.~(\ref{dive}) reduces to $\sum_i\nabla_i\hat{E}_{r,i}=0$, i.e., Gauss law without matter fields.
The second term with $\hat{E}^2_{r,i}$ is the well-known energy density of the electric field \cite{ks}. 
The third term explicitly breaks the gauge invariance under the U(1) local gauge transformation,
\be
\hat{\theta}_{r,i} &\to& \hat{\theta}'_{r,i} =\lambda_{r+i}+\hat{\theta}_{r,i} -\lambda_r, \nn
\hat{E}_{r,i} &\to& \hat{E}'_{r,i} =\hat{E}_{r,i},
\label{gt}
\ee
where $\lambda_r$ is a real $r$-dependent parameter.
However, as shown in our previous work \cite{ours1}, this term is closely related to a gauge-invariant term. 
\changedd{We will explain it later in details.}

The partition function of the quantum system $H_{\rm GH}$ of Eq.~(\ref{HGHM}) on the gauge lattice at the temperature $T$ is formulated
by the path-integral method \cite{wilson,imrev}. 
To this end, we introduce the four-dimensional lattice by piling up 3D gauge lattices along the imaginary-time ($\tau\in [0,\beta],\ 
\beta\equiv (k_{\rm B}T)^{-1}) $ direction with the spacing $\Delta \tau$.
We call this four-dimensional hypercubic lattice the (3+1)D gauge lattice, and label its sites as $x=(x_0,r)=(x_0,x_1,x_2,x_3)$, with $x_0=0,1,\cdots,L_0$ and $\Delta \tau=\beta/L_0$. 
To be precise, the limit $L_0 \to \infty$ should be taken \cite{Rothebook}.
However, in the actual Monte Carlo simulations to determine the phase structure of LGT models, it is common to keep $\Delta \tau$ finite and draw useful results by applying scaling arguments etc.~\cite{Rothebook}. 
We follow this approach in the next section.

Then the partition function of $H_{\rm GH}$ on the gauge lattice is given in the canonical formalism as follows,
\begin{eqnarray}
Z_{\rm GH}&=&
\int[DE_{x,i}][D\theta_{x,i}]\nn
&&\hspace{-0.5cm}
\times\exp\Big[\Delta\tau\Big
(i\sum_{x,i}E_{x,i}\dot{\theta}_{x,i}-\sum_{x_0}H_{\rm GH}(\theta,E)\Big)\Big],\nn
\int[DE_{x,i}]&&\hspace{-0.5cm}[D\theta_{x,i}]=\prod_{x,i}\sum_{E_{x,i}\in {\bf Z}}\int_{-\pi}^\pi
\frac{d\theta_{x,i}}{2\pi},\nn
\dot{\theta}_{x,i}&=&\frac{1}{\Delta \tau}(\changed{\theta_{x+0,i}-\theta_{x,i}})
\changed{=\frac{1}{\Delta\tau}\nabla_0\theta_{x,0},}
\label{ZGH}
\end{eqnarray}
where $H_{\rm GH}(\theta,E)$ is the \changed{c-number} obtained by replacing the operators $\hat{\theta}_{r,i}$ and $\hat{E}_{r,i}$ in $H_{\rm GH}$ of Eq.~(\ref{HGHM}) by their eigenvalues $\theta_{x,i}$ and $E_{x,i}$, respectively.

Below we follow Ref~\cite{ours1} to obtain the path integral expression of $Z_{\rm GH}$ in the Lagrange formalism in terms of the 4-component gauge field $\theta_{x,\mu} \in [0,2\pi]$ defined on the link $(x, x+\mu)$, where $\mu = 0,1,2,3$ represents the direction index (and the unit vector as before).
First, we introduce the auxiliary field $\theta_{x,0}$, the 0-th component of the vector potential, and put on the link $\changed{(x,x+0)}$ in the 0-th direction of the (3+1)D gauge lattice through the usual Gaussian integration
\be
\hspace{-0.5cm}
&&\exp\big(-\frac{\Delta \tau}{2\gamma^2}(\sum_i\nabla_i E_{x,i})^2\big)\nn
\hspace{-0.5cm}
&\propto&\int_{-\infty}^\infty d\theta_{x,0}\exp\big(-\frac{\gamma^2}{2\Delta \tau} 
\theta_{x,0}^2 +i\theta_{x,0}\sum_i\nabla_i E_{x,i}\big)\nn
\hspace{-0.5cm}
&\propto &\int_{-\pi}^\pi d\theta_{x,0}\exp\big(\frac{\gamma^2}{\Delta \tau} 
\cos\theta_{x,0}+i\theta_{x,0}\sum_i\nabla_i E_{x,i}\big).
\ee
Here, we replaced $\theta_{x,0}^2$ by $2(1-\cos\theta_{x,0})$, respecting the periodicity under $\theta_{x,0}\to \theta_{x,0}+2\pi$, which is required by the $E_{x,i}$-summation over $E_{x,i} \in {\bf Z}$. 
\changedd{The partition function $Z_{GH}$ is then given by
\be
\hspace{-0.4cm}
Z_{\rm GH}&=&
\int[DE_{x,i}][D\theta_{x,i}][D\theta_{x,0}]\exp[A(E,\theta)],\nn
\hspace{-0.4cm}
A(E,\theta)&=&
\sum_{x}\Big[ \frac{\gamma^2}{\Delta \tau}
\cos\theta_{x,0}+i\sum_i E_{x,i}\big(\nabla_0\theta_{x,i}-\nabla_i\theta_{x,0}\big)\nn
\hspace{-0.4cm}
&&-\Delta\tau\frac{V_0'}{2}\sum_{i}E_{x,i}^2
-\Delta\tau H_3\Big],\nn
\hspace{-0.4cm}
H_3&=&-\rho_{0}J\sum_{i<j}\big[\cos(\theta_{x,i}-\theta_{x,j})+
\cos(\theta_{x,i}+\theta_{x+i,j})\nn
\hspace{-0.4cm}
&&+\cos(\theta_{x+i,j}-\theta_{x+j,i})
+\cos(\theta_{x,j}+\theta_{x+j,i})
\big].
\label{ZGH3}
\ee
}

\changed{
At this stage, we derive the Gauss law which describes Eq.~(\ref{dive}) 
in a precise manner as the  Schwinger-Dyson equation \cite{peskin}
corresponding to the variation with respect to $\theta_{x,0}$.
In fact, we have the identity that the surface term in the 
$\theta_{x,0}$-integration vanishes; 
\be
&&\int[DE_{x,i}][D\theta_{x,i}][D\theta_{x,0}]\frac{\partial}{\partial\theta_{x,0}}
\exp[A(E,\theta)]=0\nn
&\to&\int[DE_{x,i}][D\theta_{x,i}][D\theta_{x,0}]\frac{\partial A(E,\theta)}{\partial\theta_{x,0}}
\exp[A(E,\theta)]=0\nn
&\to&\left\la \frac{\partial A(E,\theta)}{\partial\theta_{x,0}}\right\ra=0.
\label{SD}
\ee
This gives rise to 
\be
&&\left\la\frac{\partial}{\partial \theta_{x,0}}
\left[i\frac{\gamma^2 \cos\theta_{x,0}}{\Delta t} +i\theta_{x,0}\sum_i\nabla_i E_{x,i}\right]
\right\ra=0\nn
&\to&\sum_i\nabla_i \la E_{x,i}\ra=\la J_{x,0}\ra,\quad J_{x,0}\equiv \frac{\gamma^2\sin\theta_{x,0}}{\Delta t},
\label{dive2}
\end{eqnarray}
where $\Delta \tau$ has been replaced by $i \Delta t$
reflecting the relation for the real time $t =-i\tau$.
Eq.~(\ref{dive2}) describes the Gauss law for the present gauge Higgs model of 
Eq.~(\ref{ZGH3}) \cite{gauss5}.
}

\changed{Returning to the path to obtain the gauge-Higgs model,} 
we perform $\int [DE_{x,i}]$ for each $x$ by using Poisson's 
summation formula as
\be
\hspace{-0.3cm}
&&\prod_{x,i}\sum_{E_{x,i}}\exp\big[i\sum_{x,i}E_{x,i}
\changed{(\nabla_0\theta_{x,i}-\nabla_i\theta_{x,0})}
-\sum_{x,i}\Delta\tau\frac{V_0'}{2}E_{x,i}^2\big]\nn
\hspace{-0.3cm}&=&\prod_{x,i}\int_{-\infty}^\infty \frac{d\chi_{x,i}}{2\pi}\sum_{m_{x,i}=-\infty}^\infty
\exp\big[i\sum_{x,i}\chi_{x,i}\changed{(\nabla_0\theta_{x,i}-\nabla_i\theta_{x,0})}\nn
\hspace{-0.3cm}&&-\sum_{x,i}(\Delta\tau\frac{V_0'}{2}\chi_{x,i}^2-2\pi i\chi_{x,i} m_{x,i})\big]\nn
\hspace{-0.3cm}&\propto& \prod_{x,i}\sum_{m_{x,i}}
\exp\big[-\sum_{x,i}\frac{1}{2\Delta\tau V_0'}(\changed{\nabla_0\theta_{x,i}-\nabla_i\theta_{x,0}}+ 2\pi m_{x,i})^2\big].\nn
\label{zelecric}
\hspace{-0.3cm}\ee
The last line is just a periodic Gaussian form.
We utilize Villain's approximation in the inverse way to replace it
by the following cosine form up to a multiplicative constant; 
\be 
\exp\big[\sum_{x,i}\frac{1}{\Delta\tau V_0'}\cos\big(\changed{\nabla_0\theta_{x,i}-\nabla_i\theta_{x,0}}\big)\big].
\label{zcos}
\ee
$Z_{\rm GH}$ is then expressed as follows;
\begin{eqnarray}
\hspace{-0.7cm}
Z_{\rm GH}&=&\int [D\theta_{x,\mu}]\exp(A_{\rm GH}),\nn
A_{\rm GH}&=&A_{\rm I}+A_{\rm P}+A_{\rm L},  \nn
A_{\rm I}&=&\sum_{x,\mu}c_{1\mu}\cos \theta_{x,\mu}, \quad
A_{\rm P}=\sum_{x,\mu<\nu}c_{2\mu\nu}\cos \theta_{x,\mu\nu},  \nn
\theta_{x,\mu\nu}&\equiv&\theta_{x,\mu}+\theta_{x+\mu,\nu}
       -\theta_{x+\nu,\mu}-\theta_{x,\nu}, \nn
A_{\rm L}&=&\sum_{x,i<j}c_{3ij}
\left[\cos(\theta_{x,i}-\theta_{x,j})+\cos(\theta_{x,i}+\theta_{x+i,j})\right.\nn
&&\!\!\left.+\cos(\theta_{x+i,j}-\theta_{x+j,i})+\cos(\theta_{x,j}+\theta_{x+j,i})\right],
\label{4DGHM}
\end{eqnarray}
where $\nu = 0,1,2,3$ and $i, j =1,2,3$ are spatial indices as before.  

From $H_{\rm GH}$ in Eq.~(\ref{HGHM}) and through the way to introduce the 
scalar potential $\theta_{x,0}$, the {\em non-vanishing coefficients} in Eq.~(\ref{4DGHM}) are listed as follows;
\begin{eqnarray}
&&c_1\equiv c_{10}={\gamma^2 \over \Delta\tau},  \nonumber \\
&&c_2 \equiv c_{201}=c_{202}=c_{203}={1 \over \Delta\tau V'_0},  \nonumber \\
&&c_3\equiv c_{312}=c_{313}=c_{323}=J\rho_0\Delta\tau.
\label{cs} 
\end{eqnarray}
Therefore, the effective action $A_{\rm GH}(\theta_{x,\mu})$ has asymmetric
couplings concerning the space-time directions. 
This is in strong contrast with the LGT models of high-energy physics \cite{wilson}, which is made obvious in the plaquette term $A_{\rm P}$.
The Wilson model \cite{wilson} has $c_{2\mu\nu}=c_2$, while Eq.~(\ref{cs}) shows that $c_{2ij}=0$ for the plaquettes in the space-space directions.
Here, we note that there are some proposals for generating the space-space plaquette interactions in the context of quantum simulation of LGT using cold atomic systems \cite{Zohar1,Zohar2,Buchler,Tewari} such as making use of the second-order perturbation theory and assuming that the L-shaped interaction in $A_{\rm L}$ is a small perturbation.

We comment here on the gauge invariance in the Lagrange formalism.
The terms $A_{\rm I}$ and $A_{\rm L}$ in the effective action $A_{\rm GH}(\theta)$ break the gauge invariance under the 4D transformation, 
$\theta_{x,\mu}\to \lambda_{x+\mu}+\theta_{x,\mu}-\lambda_x$.
However, one may introduce the gauge-invariant gauge-Higgs action
$\tilde{A}_{\rm GH}(\theta,\varphi)$ for $\theta_{x,\mu}$ and the phase $\varphi_x$ 
of the Higgs field $\phi_x = \exp(i\varphi_x)$ in the London limit. This $\tilde{A}_{\rm GH}(\theta,\varphi)$ 
is {\it defined simply by the replacement} $\theta_{x,\mu} \to \theta_{x,\mu}-\varphi_{x+\mu}+\varphi_x$ {\it in} $A_{\rm GH}(\theta)$ as $\tilde{A}_{\rm GH}(\theta,\varphi)\equiv A_{\rm GH}(\theta-\varphi+\varphi)$ (its explicit form is given in Eq.~(11) of Ref.~\cite{ours1}).
It is invariant under the combined gauge transformation,
\be
\theta_{x,\mu}\to \lambda_{x+\mu}+\theta_{x,\mu}-\lambda_x, \quad
\varphi_x\to \varphi_x+\lambda_x,
\ee
by construction.
Then $A_{\rm GH}(\theta)$ becomes just the gauge fixed version of 
$\tilde{A}_{\rm GH}(\theta,\varphi)$ in
the unitary gauge $\varphi_x=0$; $A_{\rm GH}(\theta)=\tilde{A}_{\rm GH}(\theta,0)$.  
At the level of the partition function, the equivalence
\be
\tilde{Z}_{\rm GH}\equiv\int [D\theta_{x,\mu}][D\varphi_x]\exp(\tilde{A}_{\rm GH}(\theta,\varphi) )
=Z_{\rm GH},
\label{ZGHinv}
\ee 
holds.

\changed{
From this gauge-invariant
action $\tilde{A}_{\rm GH}(\theta,\varphi_x)$, some important properties 
of the Higgs field are drawn. 
The $\tilde{A}_{\rm I}$ term with $c_{1\mu} = c_1 \delta_{\mu,0}$ reads as 
\be
\tilde{A}_{\rm I} (\theta, \varphi) = \frac{c_{1}}{2} \sum_x [\phi_{x+0}^*\exp(i\theta_{x,0})\phi_x+\mbox{c.c.}].
\label{0direction}
\ee 
Its first term describes propagation of a Higgs particle
along the time axis in the positive direction as for an ordinary particle, 
whereas the second term describes its back propagation as for an antiparticle \cite{imrev}.
Therefore, the present Higgs particles are accompanied with 
their antiparticles having charges with the opposite sign. 
This is not strange at all,
although we treated the atoms described by the extended Bose-Hubbard model 
as totally nonrelativistic ones without their antiparticles. 
The Higgs charge density
is calculated from the action $\tilde{A}_{\rm I}$ in Eq.~(\ref{0direction}) 
as $\tilde{J}_{x,0}=-\partial \tilde{A}_{\rm I}(\theta,\phi)/\partial \theta_{x,0}= c_1
\sin(\theta_{x,0}-\varphi_{x+0}+\varphi_x)$, which is nothing but the 
gauge-invariant version $J_{x,0}$ appearing in Eq.~(\ref{dive2}).
This confirms that Eq.~(\ref{dive2}) represents the Gauss-law constraint
of the gauge-Higgs model defined by $\tilde{A}_{\rm GH}$.
In a similar manner, the Higgs current density $\tilde{J}_{x,i}$
is calculated from the 
$c_3$-term of the action $\tilde{A}_{\rm GH}(\theta,\phi)$ 
 as $\tilde{J}_{x,i}=-\partial\tilde{A}_{\rm L}(\theta,\phi)/\partial\theta_{x,i}
 =\sum_j c_{3ij}\sin(\theta_{x,i}-\theta_{x,j})+\cdots$.
The conservation law $\sum_\mu\nabla_\mu \tilde{J}_{x,\mu}=0$ holds owing to
the gauge invariance.
}

\changedd{
Finally, to explain our treatment of gauge invariance and the London limit explicitly, 
we  
derive $\tilde{Z}_{\rm GH}$ of Eq. (\ref{ZGHinv}) starting with the 
gauge invariant Hamiltonian, 
\be
\hspace{-0.5cm}
\hat{H}'&=&\sum_r \hat{\Pi}^\dag_r\hat{\Pi}_r+ \frac{V_0'}{2}\sum_{r,i}\hat{E}_{r,i}^2
\nn
&&-\frac{{\rho}'_{0}J}{2}\sum_{r,i<j}
\left[\hat{\Phi}^\dagger_{r+i+j}\hat{U}_{r+i,j}\hat{U}_{r,i}\hat{\Phi}_r
+ {\rm H.c.}+\cdots\right].
\ee 
Here
$\hat{\Phi}_r$ is the genuine complex Higgs field
(having radial fluctuations) and $\hat{\Pi}_r$ is its conjugate momentum
satisfying $[\hat{\Phi}_r,\hat{\Pi}_{r'}]=i\delta_{rr'}$,
and $\hat{U}_{r,i} \equiv \exp(i\hat{\theta}_{r,i})$. 
We restrict our physical space 
$|{\rm phys}\ra$ as 
\be
\hspace{-0.8cm}
&&\hat{P}|{\rm phys}\ra=0,\ \hat{P}\equiv \prod_r \delta_{\hat{G}_r, 0},\ 
\hat{G}_r \equiv \sum_i \nabla_i\hat{E}_{r,i}-\hat{Q}_r,\nn
&&\hat{Q}_r=-i:\hat{\Phi}_r\hat{\Pi}_r-\hat{\Pi}^\dag_r\hat{\Phi}^\dag_r:
=\hat{a}^\dag_r\hat{a}_r-\hat{b}^\dag_r\hat{b}_r,
\label{constraint}
\ee
where $\hat{a}$ ($\hat{b}$) is the annihilation operator of the Higgs (anti)particles,
which are introduced as $\hat{\Phi}_r \equiv (\hat{a}_r+\hat{b}_r^\dag)/\sqrt{2},\
\hat{\Pi}_r \equiv i(\hat{a}^\dag_r-\hat{b}_r)/\sqrt{2}]$.
By starting with $Z'={\rm Tr} \hat{P}\exp(-\beta \hat{H}')$, 
and following the standard method \cite{imrev2, FS}
we obtain
\be
&&Z' =\int [D\theta_{x,\mu}][D\Phi_x]\exp[A'(\theta,\Phi)],\nn
&&A'(\theta,\Phi)=\exp\left[-\sum_x\big[
\bar{\Phi}_{x+0}(\Phi_{x+0}-U_{x,0}\Phi_x)+{\rm c.c.}\big]
+A_{\rm P}\right.\nn
&&\  +\frac{{\rho}'_{0}J}{2}\sum_{x,i<j}
\left(\bar{\Phi}_{x+i+j}U_{x+i,j}U_{x,i}\Phi_x
+ {\rm c.c.}+\cdots\right),\Bigg]
\ee
where $A_{\rm P}$ is given in Eq.~(\ref{4DGHM}).
By replacing $\Phi_x$
as $\Phi_x\to \gamma/\sqrt{2\Delta\tau}\times \phi_x$ and choosing as
$\rho_0' \gamma^2/(2\Delta\tau)=\rho_0$, we obtain $\tilde{Z}_{\rm GH}$ of Eq.~(\ref{ZGHinv}).
This implies that $\tilde{Z}_{\rm GH}$ restricts the physical states
in the hard-constraint level of Eq.~(\ref{constraint}).  
As explained, by setting $\phi_x=1 (\varphi_x=0)$ we arrive at $Z_{\rm GH}$ of Eq.~(\ref{4DGHM}).
}


\section{Phase diagram of the U(1) gauge-Higgs model: MC simulation} \label{phasedia}

In the previous section, we explained how the 3D U(1) gauge-Higgs model of Eq.~(\ref{HGHM}), or equivalently its path-integral expression Eq.~(\ref{4DGHM}) on the (3+1)D lattice, appears from the extended Bose-Hubbard model of Eq.~(\ref{HEBH}) on the 3D optical lattice as its low-energy effective model.
Therefore, we expect that various dynamical properties of this gauge-Higgs model will be ``quantum simulated" by cold atomic gases in near future.
On the other hand, its static properties such as the phase structure and correlation functions may be studied by various conventional techniques. Such information is certainly useful in understanding the model and also as a guide to perform cold-atomic experiments.

In this section, we study the phase diagram of the 3D gauge-Higgs model by applying the standard (``classical") MC simulation to the (3+1)D system of 
Eq.~(\ref{4DGHM}). 
This brings no difficulties such as the negative-sign problem because the system involves only bosonic variables and has a positive definite probability.
In high-energy physics, the (3+1)D U(1) gauge-Higgs model \cite{wilson,Rothebook} that is related to the present gauge-Higgs model has symmetric couplings and is defined by Eq.~(\ref{4DGHM}) by setting $c_{1\mu}=c_1, c_{2\mu\nu}=c_2, c_{3ij}=0$. 
Its phase diagram is known to have three phases, i.e., the confinement, Coulomb, and Higgs phases \cite{wilson,imrev}. 
They are distinguished by the strength of fluctuations of $\theta_{x,\mu}$ as large, medium, and small, respectively. 
In addition, the potential energy $V(r)$ between two point sources of opposite charge and separated by distance $r$ has different typical behavior as $V(r) \propto r,\ 1/r,\ \exp(-m r)/r$, respectively.

Generally speaking, these three phases are distinguished 
by the fluctuations of the gauge field $\Delta \theta_{x,\mu}$ and 
fluctuations of the phase of the Higgs field $\Delta \varphi_x$.
The confinement phase has a large $\Delta \theta_{x,\mu}$ and $\Delta \varphi_x$, the Coulomb phase has a small $\Delta \theta_{x,\mu}$ and large $\Delta \varphi_x$, and the Higgs phase has a small $\Delta \theta_{x,\mu}$ and $\Delta \varphi_x$.
In a gauge-fixed representation such as Eq.~(\ref{4DGHM}), $\Delta \varphi_x$ is not defined. 
In this case, $\Delta \theta_{x,\mu}$ decreases in the order of confinement, Coulomb, and Higgs phases. 
In addition, one may measure the averages and fluctuations of the Higgs-coupling terms such as  $A_{\rm I}$ and $A_{\rm L}$ of Eq.~(\ref{4DGHM}) term by term. 
These values may be used to judge whether the system is in the Higgs phase or not. 

Because the present gauge-Higgs model has the asymmetric couplings $c_{1\mu}$ and $c_{2\mu\nu}$ in four directions and additional $c_{3ij}$ couplings as shown in Eq.~(\ref{cs}), its phase diagram should be examined separately, and we expect some richer phase structure.

\begin{figure}[t]
\centering
\includegraphics[width=8cm]{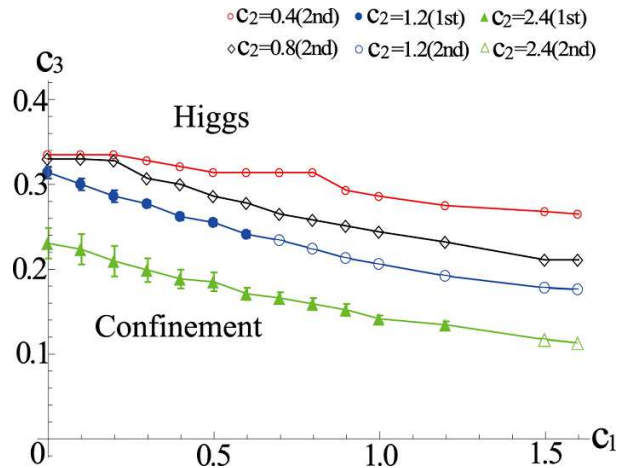}
\caption{(Color online) Phase diagram in the $c_1$-$c_3$ plane for various values of $c_2$ calculated by Monte Carlo simulation of the (3+1)D lattice 
of the size $L^4 (L=16)$.
The Higgs and confinement phases are separated by the first-order (1st) or second-order (2nd) phase-transition lines. 
The 1st and 2nd order transitions are represented by filled and empty symbols, respectively.
The transition points are measured from the peak of $C$ for a 2nd-order transition and the midpoint of the hysteresis curve of $U$ for a 1st-order transition. 
The error bars for the 1st-order transition indicate the size (starting and ending points) of hysteresis along the $c_3$ axis.
}
\label{PD}
\end{figure}

To calculate the phase diagram, we measure the following ``internal energy" $U$ and ``specific heat" $C$ as functions of the coupling constants;
\begin{eqnarray}
&& U=\langle A_{\rm GH} \rangle /L^4,  \nonumber  \\
&& C=\langle (A_{\rm GH}-\la A_{\rm GH}\ra)^2 \rangle/L^4,
\label{UC}
\end{eqnarray}
where we consider the (3+1)D space-time hypercubic lattice with the common linear size $L_\mu$ in the $\mu$-th direction, and use $L_\mu=L$ with periodic boundary condition \changed{\cite{asymlattice}}.
An explanation of our MC calculations and some supplementary results such as scaling analysis are given in Appendix \ref{mc}. 
The thermodynamic limit is given by taking $L\to \infty$, which we shall discuss later, together with the path-integral requirement $\Delta \tau \to 0$.
We determine the order of the phase transition by checking the behaviors of 
$U$ and $C$ as follows; (i) If $U$ exhibits a hysteresis (a jump $\Delta U \neq 0$) as we change a parameter back and forth, it is a first-order transition; (ii) If $C$ has a peak increasing as $L$ increases, it is a second-order transition; (iii) If $U$ has no hysteresis and the peak is round or does not develop as $L$ increases, it is a crossover (no genuine 
transition) \cite{kt}.

\begin{figure}[t]
\centering
\hspace{-0.3cm}
\includegraphics[width=8.5cm]{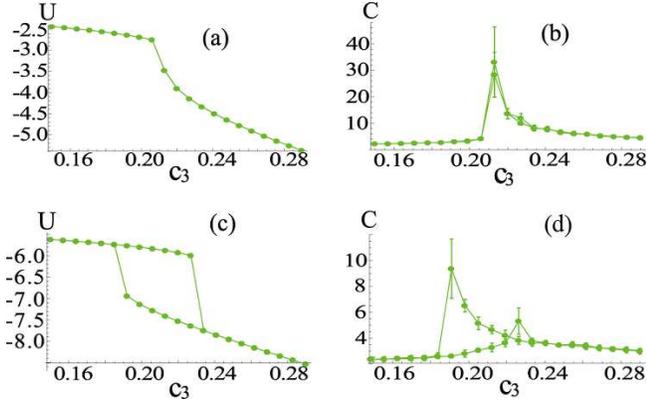}
\hspace{-0.3cm}
\caption{(Color online) Typical behavior of the internal energy $U$ and 
specific heat $C$ of Eq.~(\ref{UC}) for $L=16$.
(a) $U$ and (b) $C$ for $c_1=0.9, c_2=1.2$ show a second-order phase transition at $c_3\simeq 0.21$. 
(c) $U$ and (d) $C$ for $c_1=0.2, c_2=2.4$ show a first-order phase transition at $c_3\simeq 0.185\sim 0.235$.
The hysteresis loop in (c) is obtained as we first increase $c_3$ and then decrease it.
}
\label{UC1}
\end{figure}

In Fig.~\ref{PD}, we show the phase diagram in the $c_1$-$c_3$ plane for
several fixed values of $c_2$.
There are two phases. 
Below we shall see that the phase in the lower $c_3$ region is the confinement phase and the phase in the higher $c_3$ region is the Higgs phase. 
They are separated by a first-order or second-order phase transition line.
In Fig.~\ref{UC1}, we present the two sets $U$ and $C$ as functions of $c_3$ for fixed $c_1 and c_2$, which exhibit a typical second-order 
and first-order transition, respectively.
For the first-order transitions, the hysteresis effect obscures the location of the transition point.
In Fig.~\ref{PD}, we plot the midpoint of the hysteresis curve as the transition point.
For a precise determination of the transition point, we need another algorithm such as the multicanonical ensemble \cite{multicanonical}, which is a future problem.

To identify the nature of the two phases in Fig.~\ref{PD}, we measure the fluctuations (uncertainty) $\Delta E$ of the electric field $\vec{E}$ and the fluctuation $\Delta B$ of the magnetic field $\vec{B}$. 
Explicitly we use the following quantities;
\begin{eqnarray}
W_{\rm e}&\equiv&{1 \over 3L^4}\sum_{x,i}\la (E_{x,i}-\la E_{x,i}\ra)^2\ra\nn
&=&{1 \over 3L^4}\sum_{x, i}
\biggl[c_{2}\la\cos\theta_{x, 0i}\ra -c^{2}_{2}\la\sin^{2}\theta_{x,0i}\rangle\biggr],
\label{we}\\
W_{\rm m}&\equiv&{1 \over 3L^4}\sum_{x, i<j}
\la(\sin\theta_{x,ij}-\la\sin\theta_{x,ij}\ra)^2\ra,
\label{wm}
\end{eqnarray}
where $\theta_{x,\mu\nu}$ is defined in Eq.~(\ref{4DGHM}), i.e., 
the field strength defined on the plaquette $(x,x+\mu,x+\mu+\nu,x+\nu)$.
In Eq.~(\ref{we}), the second equality is obtained by following the path from Eqs.~(\ref{ZGH}) to (\ref{4DGHM}) by adding the source term for $E_{x,i}$ to the action. 
The detailed derivation of $W_{\rm  e}$ is shown in Appendix \ref{electricterm_Eq18}.
Because the relation $B_{i}(x)\propto \sum_{jk}\epsilon_{ijk}\theta_{x,jk} (\epsilon_{ijk}$ is the completely anti-symmetric tensor) holds in the continuum limit \cite{wilson,ks,Rothebook}, its simple compactification $\sum_{jk}\epsilon_{ijk}\sin\theta_{x,ij}$ is taken as a natural candidate for the magnetic field $B_{x, i}$ on lattice \cite{polyakov,mddef}. 
Then $W_{\rm m}$ of Eq.~(\ref{wm}) is just the square of the fluctuation, $(\Delta B)^2 =(3L^4)^{-1} \sum_{x,i}\la (B_{x,i} -\la B_{x,i}\ra)^2\ra$.

\changed{
There exist the correlations that large $W_{\rm m}$ implies large $\Delta B$ and
$\Delta \theta_{x,i}$, and large $W_{\rm  e}$ implies large $\Delta E$ and {\it small} $\Delta \theta_{x,0}$ because $\theta_{x,0}$ is the variable  conjugate to
$\sum_i\nabla_iE_{x,i}$.
}
From the characterization of each phase given above, the confinement phase has small $\Delta E$ and large $\Delta B$, while the Higgs phase and Coulomb phase have small $\Delta B$ and large $\Delta E$.  

\begin{figure}[t]
\centering
\hspace{-0.3cm}
\includegraphics[width=8.5cm]{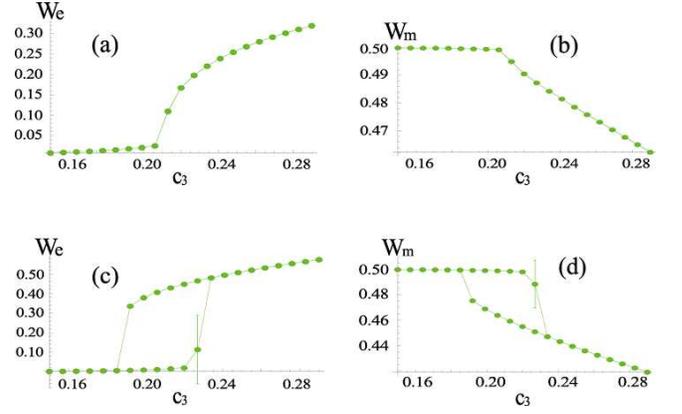}
\hspace{-0.3cm}
\caption{(Color online) Fluctuation strengths of the electric and magnetic fields, $W_{\rm e}$ and $W_{\rm m}$ of Eqs.~(\ref{we}) and (\ref{wm}) for $L=16$. 
Panels (a) and (b) show $W_{\rm e}$ and $W_{\rm m}$, respectively, for $c_1=0.9$ and $c_2=1.2$, while (c) and (d) show those for $c_1=0.2$ and 
$c_2=2.4$, respectively.
$W_{\rm e}$ is small (large) in the smaller (larger) $c_3$ region, whereas $W_{\rm m}$ behaves the other way around.
From their behaviors, we can identify the confinement and Higgs phases
as in the phase diagram in Fig.~\ref{PD}.
}\label{CalWEWM}
\end{figure}

In Fig.~\ref{CalWEWM}, we show $W_{\rm e}$ and $W_{\rm m}$ for the parameters chosen in Fig.~\ref{UC1}. 
In the phase with smaller $c_3$, $W_{\rm e}$ ($W_{\rm m}$) is small (large).
Hence, this phase is the confinement phase. 
On the other hand, since $W_{\rm e}$ ($W_{\rm m}$) is large (small) in the phase with larger $c_3$, this phase can be either the Higgs or Coulomb phase.
Because larger $c_3$ implies that the $c_3$ Higgs-coupling term $A_{\rm L}$
has a larger expectation value and smaller fluctuations than the confinement phase at smaller $c_3$, it should be the Higgs phase. 
This conclusion is confirmed by measuring $\la A_{\rm L}\ra$ and the $c_3$ specific heat $(\Delta A_{\rm L})^2\equiv \la A_{\rm L}^2\ra - \la A_{\rm L}\ra^2$ directly.

\begin{figure}[t]
\centering
\hspace{-0.3cm}
\includegraphics[width=8.5cm]{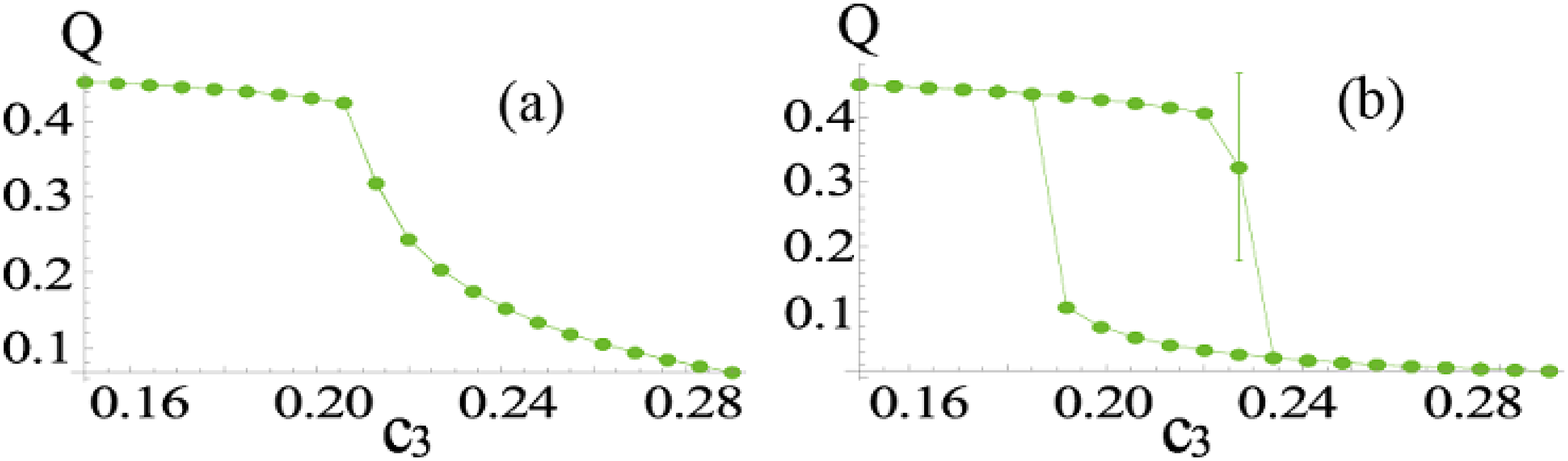}
\caption{(Color online) Magnetic monopole density $Q\equiv\la Q_x\ra$ [See Eq.~(\ref{Qx})] for $L=16$; (a) $c_1=0.9, c_2=1.2$ and b) $c_1=0.2, c_2=2.4$.
They have behaviors that are similar to the corresponding $W_{\rm m}$ in Fig.~\ref{CalWEWM}.
}
\label{md}
\end{figure}

The characteristics of the confinement phase, i.e., largeness of $\Delta B$, is sometimes rephrased as a condensation of magnetic monopoles \cite{polyakov}.
Magnetic monopoles describe topologically nontrivial configurations of the magnetic field strength $\theta_{x,ij}$, i.e., configurations having ``large" $\theta_{x,ij}$. 
To define the monopole density $Q_x$, we decompose $\theta_{x,ij}$ into its {\em integer (large) part} $2\pi n_{x,ij} (n_{x,ij} \in {\bf Z})$ and the remaining (small) part $\tilde{\theta}_{r,ij}$ as
\begin{equation}
\theta_{x,ij}=2\pi n_{x,ij}+\tilde{\theta}_{x,ij}, \;\; (-\pi <\tilde{\theta}_{r,ij}<\pi),
\end{equation}
where $n_{x,ij} (\neq 0)$ describes nothing but the Dirac string (quantized magnetic flux) penetrating the plaquette $(x,ij)$. 
Then, the monopole density $Q_x$ is defined \cite{mddef} as
\begin{eqnarray}
Q_x&\equiv&-{1 \over 2}\sum_{i,j,k}\epsilon_{ijk}(n_{x+i,j,k}-n_{x,jk})   \nonumber \\
&=&{1 \over 4\pi}\sum_{i,j,k}\epsilon_{ijk}(\tilde{\theta}_{x,jk}
-\tilde{\theta}_{x,jk}),
\label{Qx}
\end{eqnarray}
where the last equality comes from the identity $\sum_{i,j,k} \epsilon_{ijk} (\theta_{x+i,jk}-\theta_{x,jk}) = 0$ (lattice version of div$\cdot$rot $=0$).
Therefore, $Q_x$ measures the total magnetic fluxes emanating from the 6 surfaces (plaquettes) of the 3D cube centered at the dual lattice site $x+{\hat{1} \over 2}+{\hat{2} \over 2}+{\hat{3}\over 2}$.
$Q_x$ certainly expresses the magnitude of the topologically nontrivial 
fluctuations of the space-component of the gauge field $\theta_{x,i}$ in a local and gauge-invariant manner. 
In Fig.~\ref{md}, we plot the average $Q\equiv\la Q_x\ra$ for the two cases shown in Figs.~\ref{UC1} and \ref{CalWEWM}.
It has a behavior similar to that of $W_m$ of Fig.~\ref{CalWEWM}; As expected, $Q$ is large in the confinement phase and very small in the Higgs phase.

To understand the phase structure of the gauge-Higgs model, let us focus on the order of the phase transitions in Fig.~\ref{PD}.
This may be summarized as follows; as $c_3$ increases while $c_1$ and $c_2$ are fixed, the transition from the confinement phase to the Higgs phase is 
second order for large $c_1$ and small $c_2$ and shifts to first order as $c_1$ decreases and/or $c_2$ increases.
This crossover of the order is an interesting phenomenon itself.
One may conceive a few plausible arguments to explain this point.
Although it is not rigorous, we present one such an argument in Appendix \ref{interprephasedia}.
It is based on known facts about the related models and an interpretation of the system (\ref{4DGHM}) as a sum of mutually interacting two XY spin systems. 
One system has the action $A_{\rm I}$ of Eq.~(\ref{4DGHM}) and consists of the time-component of the gauge field $\theta_{x,0}$.
The other system has the action $A_{\rm L}$ of Eq.~(\ref{4DGHM}) and consists of the space-component $\theta_{x,i}$. 
A synthetic effect between these two systems, driven by the coupling action  $A_{\rm P}$ of Eq.~(\ref{4DGHM}), may convert an ordinary second-order transition to a first-order one.
This is one of the characteristics of the present system having asymmetric couplings in the space-time directions, which reflects the nonrelativistic nature of the starting extended Bose-Hubbard model. 
This is in sharp contrast with the LGT studied in high-energy physics \cite{wilson}, which has symmetric couplings in the space-time directions
reflecting the relativistic invariance.

At this point, we mention the possibility of the Coulomb phase in our system. 
A typical example of the Coulomb phase is the ordered phase of $\theta_{x,\mu}$ in the 4D Wilson model ($c_{1\mu}=c_{3\mu\nu}=0, 
c_{2\mu\nu}=c_2$) for $c_2 \gtrsim 1.0$. 
For the symmetric 4D U(1) gauge-Higgs model ($c_{1\mu}=c_1,  c_{2\mu\nu}=c_2, c_{3\mu\nu}=0$), it appears in the region of large $c_2$ and small $c_1$ (it is a smooth extension from the Coulomb phase of the Wilson model). 
It is also known that the 3D Wilson model has only the confinement phase and no Coulomb phase \cite{polyakov}. 
In our system, because the spatial-spatial plaquette term is missing, $c_{2ij}=0$, it is rather hard to expect the Coulomb phase. 
In fact, we checked that the specific heat $C(c_2)$ at $c_1=c_3=0$ has no peaks developing as $L (\leq 16)$ increases.
This indicates that the Coulomb phase does not show up and the confinement phase dominates along the $c_2$ axis.
In addition, we comment here on the approach in Ref.~\cite{Zohar2}.
It is argued there, in the context of the gauge magnet, that the second-order perturbation of a small $c_{3ij}$ term may generate the $c_{2ij}$ term effectively, but the present argument and Fig.~\ref{PD} indicate that such a $c_{2ij}$ is not large enough to generate the Coulomb phase.

Next, we discuss how to manage the limit of $\Delta \tau \to 0$, which is required in the precise path-integral treatment.
This is important when we use the present MC result such as the phase diagram Fig.~\ref{PD} as a guide to set up experiments and interpret their results.  
For practical MC simulations, as mentioned in Sec.~\ref{formulation}, we use sufficiently large but finite size $L_0$ in the imaginary-time direction with the finite-size scaling hypothesis, which, in our symmetric choice $L_\mu = L$, implies the thermodynamic limit at the same time.  
We are interested in the sufficiently low temperature region $T < T_{\rm BH}$, where $T_{\rm BH}$ may be $\sim$ 10 nK by setting the parameters of $H_{\rm EBH}$ suitably to focus on quantum phase transitions instead of thermal phase transitions. 
This temperature region $T\simeq 0$ is consistent with the limit $L\to \infty$ in our symmetric choice $L_\mu=L$ \cite{wilson,Rothebook}.

Equation~(\ref{cs}) shows that the limit $\Delta \tau \to 0$ with the physical parameters $\gamma^{-2}, V_0', J, \rho_0$ kept finite implies that the dimensionless parameters approach the limits $c_1, c_2 \to \infty,  c_3 \to 0$ in the space of $c_i$. 
To discuss the possible phase transition, etc., we need to ``enlarge" this limiting point in some way.
This is possible by calculating the transition points {\it for finite $c_i$'s } by using $U$ and $C$ at sufficiently large, but finite, $L$'s such that they exhibit scaling behaviors. 
Then we map these points into {\it a new space parameterized by $\Delta \tau$-independent combinations of $c_i$'s}, and extrapolate these boundaries to the limit $c_1, c_2 \to \infty, c_3 \to 0$.
These extrapolated boundaries are genuine transition points for $\Delta\tau\to 0$. 
They are not a single point anymore and carry nontrivial information as in a typical phase diagram in a certain parameter space for quantum phase transition.

\begin{figure}[t]
\centering
\includegraphics[width=7.5cm]{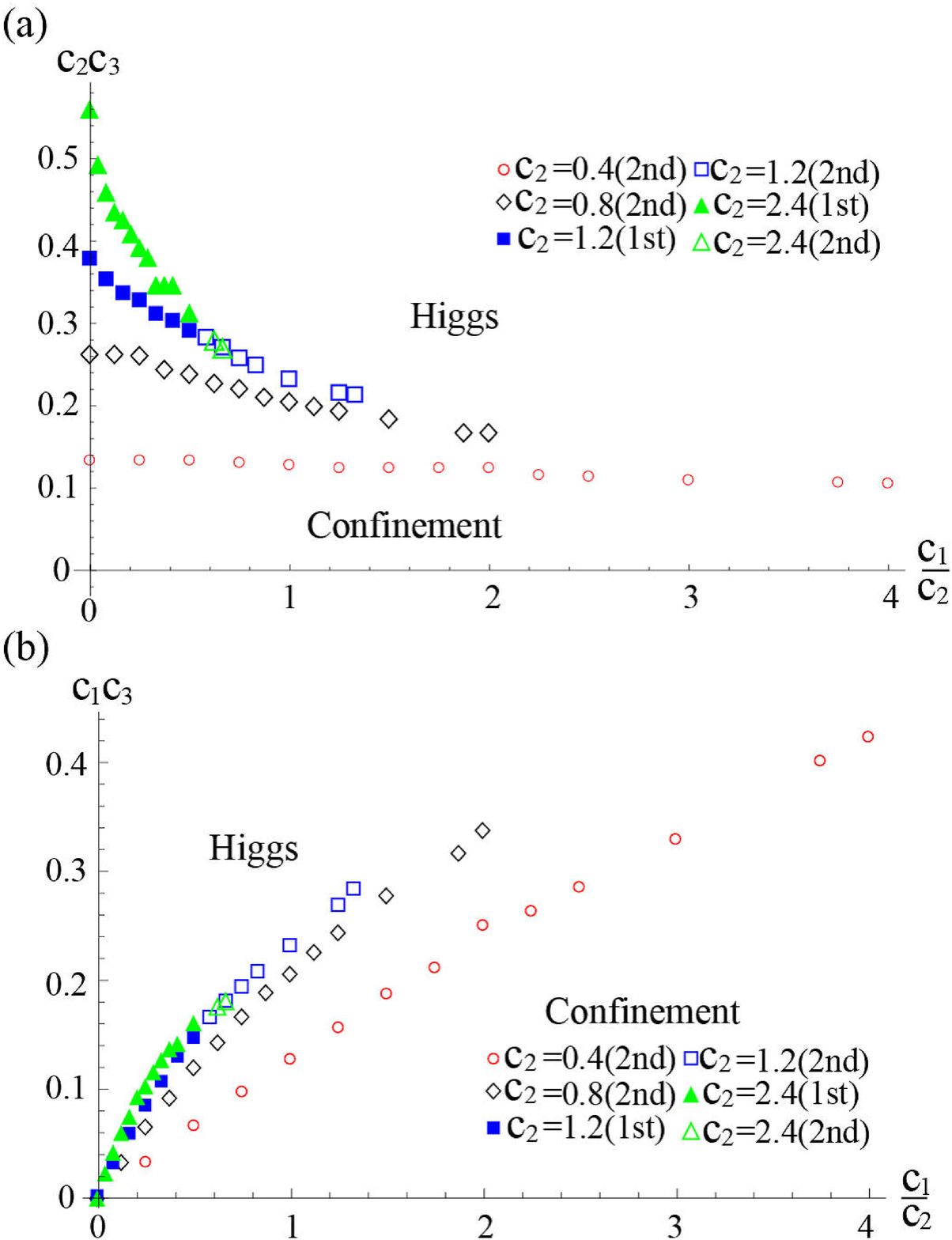}\\
\vspace{0.5cm}
\includegraphics[width=7.0cm]{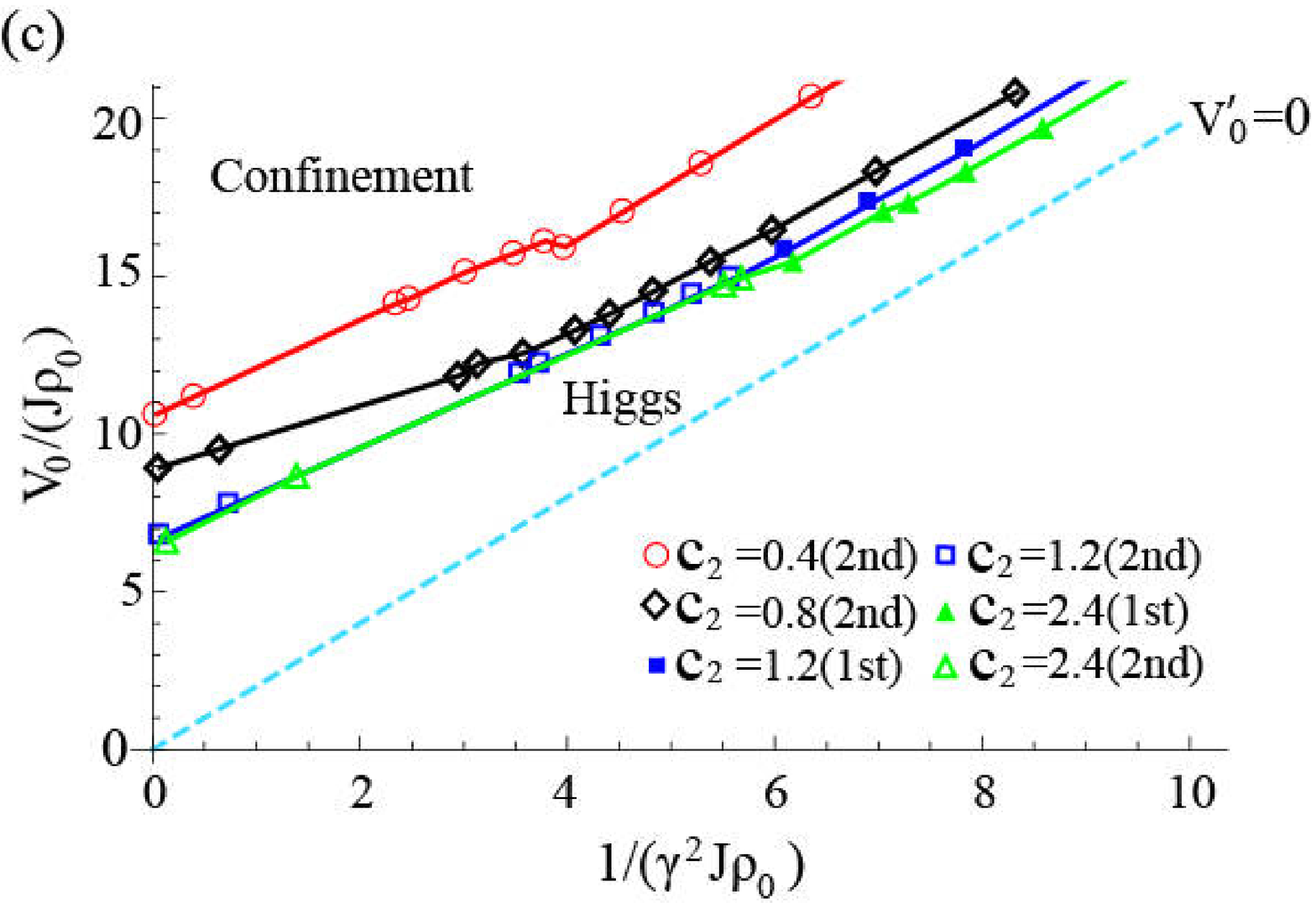}
\vspace{-0.5cm}
\caption{(Color online) Phase diagrams for various values of $c_2$, shown in Fig.~\ref{PD}, are redrawn in the planes of dimensionless coordinates; 
(a) $c_1/c_2$-$(c_2 c_3)$ plane,
(b) $c_1/c_2$-$(c_1 c_3)$ plane, and 
(c) $(c_1c_3)^{-1}$-$[(c_2c_3)^{-1}+2(c_1c_3)^{-1}]$ plane. 
In terms of the original parameters, they read (a) $\gamma^2 V_0'$-$J\rho_0/V_0'$, 
(b) $\gamma^2 V_0'$-$\gamma^2 J\rho_0$, and (c) $1/(\gamma^2J\rho_0)$-
$V_0/(J\rho_0)$. 
The 1st and 2nd order transitions are represented by filled and empty symbols, respectively.  
In Fig.~(c) the points on the vertical axis ($\gamma^2J\rho_0\to \infty$) 
are extrapolations from the nearest three points.
The dashed line in Fig.~(c) is the line of $V_0'=0$.
The present gauge Higgs model is defined only in the region $V_0' >0$ above this line (See the text).
}
\label{PD2}
\end{figure}

To follow this program explicitly, we redraw in Fig.~\ref{PD2} the four phase-boundary curves for each $c_2$ in Fig.~\ref{PD} into the two-dimensional plane of the horizontal axis; $c_1/c_2 = \gamma^2 V_0'$ and the vertical axis $c_2\cdot c_3= J\rho_0/V_0'$ [Fig.~\ref{PD2}(a)] and 
$c_1\cdot c_3= \gamma^2 J\rho_0$ [Fig.~\ref{PD2}(b)].
Here we note that Fig.~\ref{PD} is drawn by using the data of $L=16$, where $L=16$ is supported to be in the scaling region by the scaling analysis in 
Appendix B.
Then we discuss the extrapolation of these boundaries as $c_2\to\infty$
\cite{scaling0}.
We note that the dimensions of the parameter space reduce as 4 $({\rm original;}\ \gamma^2,J,\rho_0,\Delta\tau) \to $ 3 $({\rm dimensionless;}\ c_1,c_2,c_3) \to$ 2 $(\Delta \tau$-eliminated; $c_1/c_2, c_2(c_1)\cdot c_3)$.  

In Fig.~\ref{PD2}(a), the boundary curve shifts upward systematically as $c_2 (= 0.4,0.8,1.2, 2.4)$ increases, but its part of the second-order transition seems to converge as $c_2\to\infty$ to a fixed curve given by $c_2=1.2$ and 2.4 (they are almost degenerate).
Part of the first-order transition still develops and there is no sign of convergence.
In Fig.~\ref{PD2}(b), the four boundary curves show a rather clean convergence behavior to the degenerate curves of $c_1=1.2$ and 2.4, where the deviation in the first-order points above are not visible because the relevant region in Fig.~\ref{PD2}(a) is now condensed near the origin. 
In Fig.~\ref{PD2}(c), we replot the four curves of Fig.~\ref{PD} in the $1/(\gamma^2J\rho_0)$-$V_0/(J\rho_0)$ plane, i.e., the normalized (off site)-(on site) density-density interactions at large $\rho_0$. 
Similar phase diagrams are often drawn for the systems with small fillings (small $\rho_0$) \cite{ours3}, and this diagram may be helpful in choosing interaction parameters in the experimental setups. 
Again, it shows rather clear convergence behavior of the curves of $c_2=1.2$ and 2.4 \cite{temperature}. 

These observations lead us to the conclusion that the degenerate curves of $c_2=1.2$ and 2.4 in Fig.~\ref{PD2}(b) and (c) give rise to approximate but useful locations for the genuine transitions defined in the limit of $\Delta\tau\to 0$. 
In actual quantum simulations of the gauge theory, one may predict the location of a phase transition point of the gauge-Higgs model of 
Eq.~(\ref{4DGHM}) from a single transition point determined experimentally (e.g., as discussed in Sec.~\ref{gpdynamics}) in the two-dimensional plane of Fig.~\ref{PD2}(c). 
Here we note that an experimentally determined point corresponds to a point on the ``would-be" line of $c_2=\infty$ in Fig.~\ref{PD2}(c), but, of course, the correspondence is not exact due to the approximations involved in the derivation of the gauge-Higgs model in Sec.~\ref{formulation}. 
To locate the transition point in the gauge-Higgs model, one may simply choose an arbitrary value of $c_2$, {\it which should be larger than $1.2$}, and calculate the corresponding critical values of $c_3$ and $c_1$ according to the mapping rule between Figs.~\ref{PD} and \ref{PD2}(c).
This confirms that the results of Fig.~\ref{PD} are useful as long as 
$c_2 \gtrsim 1.2$.
This ``magic" in dimensional unbalance to produce a gauge-Higgs phase diagram in the 3D $c_1$-$c_2$-$c_3$ space ($c_2\gtrsim 1.2$) from the phase diagram of the extended Bose-Hubbard model in 2D $V_0/J$-$V_{ab}/J$ space is based on the quick convergence of the transition curves explained above. 
This is a nontrivial observation.

To close this section we comment on Fig.~\ref{PD2}(c). 
As Eqs.~(\ref{HGHM}) and (\ref{zelecric}) show, we assumed $V_0' > 0$ in 
deriving the present gauge-Higgs model, which implies the region above the dashed line $V_0'=0$ in Fig.~\ref{PD2}(c) \cite{negativevprime}.
As noted before, the Higgs phase of the gauge-Higgs model corresponds to 
the superfluid phase of atomic system and the confinement phase to the Mott-insulator state. 
Concerning the latter correspondence, we note that, in the Mott state realized in some region of $V_0' < 0$, the density fluctuations may be short-range. 
However, in the confinement regime, here with $V_0' > 0$, the density fluctuations may have long-range correlations because they correspond to the electric field and a one-dimensional electric flux connecting a pair of opposite charges may be formed in the confinement phase (see Sec.~\ref{gpdynamics}).
Therefore, not only the on-site interaction but also the long-range interactions between atoms must be carefully adjusted to realize the confinement phase.

\section{Dynamics of electric flux by semiclassical approximation of the gauge-Higgs model} \label{gpdynamics}

In the previous section, we obtained the phase diagram of the effective gauge system by MC simulation in Figs.~\ref{PD} and \ref{PD2}, which is one of the most important static properties of the system under question.
In this section, we study the dynamical properties of each phase in this
phase diagram; in particular, we are interested in the behavior of an electric flux connecting a pair of external charges. 

It is a challenging problem to explore real-time dynamics of quantum many-body systems. 
Recently, a tensor network method was applied to study the real time 
dynamics of string breaking for a (1+1)D quantum link model \cite{Pichler}, 
but this method is restricted to problems in one spatial dimension. 
Here, we use a simple mean field treatment to study the real time dynamics 
by following our previous studies \cite{ours2,ours3}, in which the semiclassical GPE (discrete nonlinear Schr\"{o}dinger equation) was employed to study the dynamics of an electric flux in the gauge-Higgs model derived from the atomic system. 
We note that the direct derivation and application of GPE for gauge theories, including LGT, is not straightforward due to the existence of gauge invariance; one needs to determine how to respect gauge invariance. 
The correspondence between the extended Bose-Hubbard and the gauge-Higgs model we have explained so far offers us a convincing approach; one can derive and solve the GPE of the extended Bose-Hubbard model and use the correspondence between the two sets of parameters to interpret the solution
in the context of gauge theory. 
\changed{
Here we should mention that the analysis using the GPE corresponds
to the LGT with the temporal system size $L_0$ much larger than
the spatial lattice size $L_i$.
}

The Gross-Pitaevskii description of the extended Bose-Hubbard model may be quite effective if the site occupation is large enough and the phase coherence in each site is well established, because the field operator at each site is just replaced by the $c$-number field within the GPE \cite{latticegpe}. 
As explained before, this regime is within our assumptions for realizing the gauge-Higgs model. 
More precisely, the Higgs phase can be described well by GPE, because 
it corresponds to the superfluid phase. 
As we approach the confinement phase, GPE cannot work well to study the dynamics, because quantum fluctuation becomes large. 
However, we still expect that some qualitative feature can be captured by the Gross-Pitaevskii approach, as described in our previous studies \cite{ours2,ours3} and later discussion. 
Quantum fluctuations can be included in the truncated Wigner approximation \cite{TWA}, which is obtained by taking into account quantum fluctuations around the classical path up to the second-order. 
The truncated Wigner approximation consists of (i) deriving an equation of motion of the average value of quantum operator, which is just the GPE itself;  (ii) solving GPE for a given initial condition; and (iii) 
averaging over the solutions of GPE with different initial conditions
with a certain weight. 
The faithful treatment according to the truncated Wigner approximation requires step (iii), which certainly seems important because $\hat{E}_{r,i}$ and $\hat{\theta}_{r,i}$ are canonically conjugate pairs and their averages should obey the uncertainty principle. 
The implementation of the requirement of step (iii) into actual experiments and the discussion of the appropriateness of the result with a single initial condition is discussed quantitatively for gauge-Higgs model in one spatial dimension \cite{ours3}. 
We leave this discussion for the present 3D model as a future problem, and focus on the detailed study with the most interesting initial condition below, which is certainly important by itself.

The equation of motion for the effective gauge model of Eq.~(\ref{HGHM}) 
involves the expectation values $E_{r,i}(t)$ and $\theta_{r,i}(t)$ ($t$ is the time) of the operators $\hat{E}_{r,i}$ and $\hat{\theta}_{r,i}$ of Eq.~(\ref{AE}), respectively.
It may be obtained by averaging the Heisenberg equations of motion for $\hat{E}_{r,i}$ and $\hat{\theta}_{r,i}$ and truncating the quantum correlations among them or by taking the saddle point configuration of the path integral in canonical formalism (\ref{ZGH}).
Explicitly, we have
\begin{widetext}
\begin{eqnarray}
&&\hbar{d\over dt}E_{r,i}=-2J\rho_{0}\sum_{j = 1, 2, 3 (\neq i)}
[\sin(\theta_{r,i}-\theta_{r,j})+\sin(\theta_{r,i}+\theta_{r-j,j})
+\sin(\theta_{r,i}+\theta_{r+i,j})+\sin(\theta_{r,i}-\theta_{r+j-j,j})] \nonumber \\
&&\hbar{d \over dt}\theta_{r,i}=V'_0 E_{r,i}
+{1 \over \gamma^2}\Big[E_{r,i}-E_{r-i,i}
+\sum_{j=1,2,3 (\neq i)}(E_{r,j}-E_{r-j,j})\Big] \nonumber \\
&&\hspace{2cm}+{1 \over \gamma^2}\Big[-E_{r+i,i}+E_{r,i}
+\sum_{j=1,2,3(\neq i)}(-E_{r+i,j}+E_{r+i-j,j})\Big],
\label{GPE}
\end{eqnarray}
\end{widetext}
for a canonically conjugate pair $\theta_{r,i}$ and $E_{r,i}$. 
We are interested in the motion of an electric flux initially pinned up between two external static source charges. 
Therefore, we confine ourselves to the solution of Eq.~(\ref{GPE}) with the initial condition representing such a situation.

We solve Eq.~(\ref{GPE}) by the standard Crank-Nicolson method with a discrete time step $\Delta t$.
We use a 3D cubic lattice (gauge lattice) with the size $100\times 100 \times 100$, i.e., we define the lattice site $r=(r_{x},r_{y},r_{z})$ with $1\leq r_{i}\leq 100$, and apply the Neumann boundary condition. 
Concerning the dimensionless time step, we use $\Delta\tilde{t}\equiv V_0'\Delta t/\hbar$ and set $\Delta\tilde{t} = 0.01$, and then make runs with typical elapsed time steps $20000\sim 30000 (\times \Delta\tilde{t} )$.
The realistic time scale corresponding to this choice can be estimated as
$\Delta t = 0.01\times \hbar/V_0' \sim$ 0.0032 ms for the typical energy scale $V'_{0}/h\sim$ 500 Hz used in experiments \cite{Endres}.
For other dimensionless parameters, we considered the case of 
$\tilde{\gamma}^2\equiv \gamma^2V_0'=1$ and $10$, and $\tilde{J}\equiv J\rho_0 /V'_0=0.001-10$.
These parameters correspond to $c_1/c_2=\tilde{\gamma}^2$ and $c_2\cdot c_3 =\tilde{J}$ in Fig.~\ref{PD2}(a). 
For example, as $\tilde{J}$ increases from 0.001 to 10 for $\tilde{\gamma}^2=1$, the system moves from the deep confinement region to the deep Higgs region in the phase diagram of Fig.~\ref{PD2}(a). 

\begin{figure}[t]
\centering
\includegraphics[width=8cm]{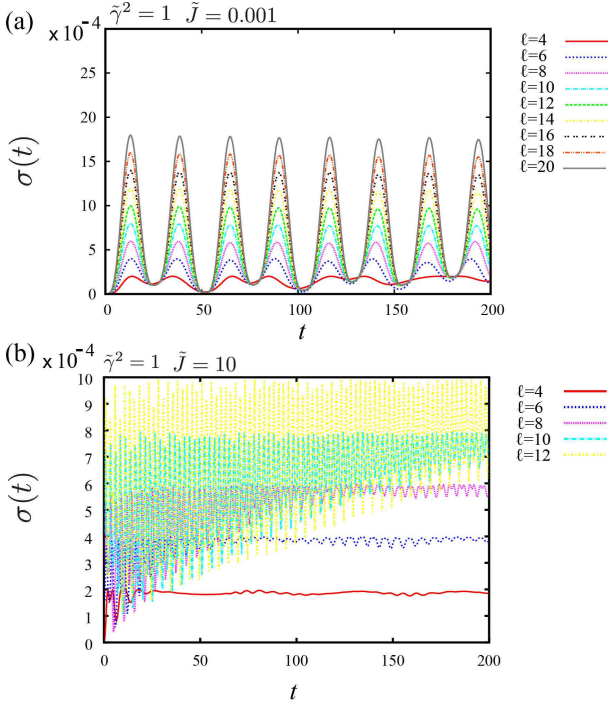}
\caption{(Color online) Time evolution of $\sigma(t)$ of Eq.~(\ref{sigma})
for various lengths $\ell$ of electric flux with $E_0=0.1$. 
The amplitude of the oscillation of $\sigma(t)$ increases with the 
length $\ell$ according to the definition of Eq.~(\ref{sigma}).
(a) $\tilde{J}(\equiv J\rho_0 /V'_0)=0.001$ (confinement);
$\sigma(t)$ continues oscillations and shows the stability of the electric flux.
(b) $\tilde{J}=10$ (Higgs); the oscillation of $\sigma(t)$ gradually diminishes and $\sigma(t)$ approaches an $\ell$-dependent constant $\simeq (\ell -2) E_0^4$.
} 
\label{sigma_t}
\end{figure}

We set up our simulation by pinning down two external charges separated by the distance $\ell$; a positive charge $q_+$ at the site $r_{+}=(50-\ell/2,50,50)$ and a negative charge $q_-$ at $r_{-}=(50+\ell/2,50,50)$.
Because there are no dynamical variables for charges on the sites in Eq.~(\ref{GPE}), we fix alternately the electric field $E_{r_+,1}(t)$ emitted from $q_+$ and directed to $q_-$, and $E_{r_{-}-1,1}(t)$ absorbed on $q_-$ as $E_{r_+,1}(t)= E_{r_{-}-1,1}(t) = E_0 (>0)$ throughout the process.  
As the initial condition of Eq.~(\ref{GPE}), we prepare an electric flux of strength $E_0$ connecting $q_+$ and $q_-$ by setting $E_{r,1}(0)=E_0 $ along the straight line L$_{\pm}$ spanned between $r_+$ and $r_-$, and $E_{r,i}(0) = 0$ for other links. We set $\theta_{r,i}(0)=0$.
 
To study the stability of the electric flux, we measure fluctuations of the electric field (fluctuation of atomic density) by using the quantity: 
\begin{equation}
\sigma(t)\equiv \sum_{(r,i)\in\ {\rm L}_\pm}\biggl[(E_{r,i}(t))^2-E^2_0\biggr]^2,
\label{sigma}
\end{equation}
where the summation is taken for the sites on the straight line L$_\pm$.
If the initial flux configuration is stable for a long time, $\sigma(t)$ stays close to zero. 

In Fig.~\ref{sigma_t} (a), we show the time development of $\sigma(t)$ in the deep confinement phase ($\tilde{J}=0.001$) for various $\ell$. 
$\sigma(t)$ exhibits an approximately periodic oscillation with its amplitude having larger values for larger $\ell$. 
This oscillation in time comes from the Higgs coupling ($J\rho_0$-term) that is present even for the present case.
When one of these terms $\exp(i\theta_{r,j})\exp(-i\theta_{r,i})$ is applied to an electric flux bit at $(r,r+i)$, it annihilates this bit and creates a new bit at $(r,r+j)$ \cite{ks}. 
The next application may restore the original flux, and complete a process of vacuum polarization by creation and annihilation of a pair of Higgs particles. 
This phenomenon is closely related to the Schwinger mechanism \cite{Schwinger}, and has also been seen in previous literature \cite{ours2,Pichler,ours3}.
More generally, these oscillations exhibit just an exchange of energy between the kinetic energy ($V_0'$-electric term) and the potential energy ($J\rho_0$ Higgs term).
This can be understood by referring to {\it the case without external charges $q_{\pm}$}. 
Then the uniform configuration $E_{r,i}(t)= E(t)$ and $\theta_{r,i}(t)=\theta(t)$ has the conserved energy density $W$ per link (the same form as a simple gravity pendulum) and the approximate harmonic-oscillator solution of Eq.~(\ref{GPE}) for $|\theta(t)| \ll 1$; 
\be
W&=&  \frac{V_0'}{2}E(t)^2-4J\rho_0 \cos(2\theta(t)),\ \
E(t)=\frac{\hbar}{V_0'}\frac{d}{dt}{\theta}(t),\nn
E(t)&\simeq& \tilde{E}\cos(\omega t+ \alpha),\ \
\omega^2= \frac{16J\rho_0V_0'}{\hbar^2}.
\label{globaloscillation}
\ee 
As we shall see, the total energy of the system is not strictly conserved due to pinning down two charges.
However, the oscillation in Fig.~\ref{sigma_t}(a) is taken as a local realization of Eq.~(\ref{globaloscillation}) with the frequency $\omega$ along the line L$_\pm$.
We conclude that the electric flux oscillates but is stable for a long time for the present parameters. 
This is a clear evidence showing that the system stays in the confinement phase, as we expected from the phase diagram by the MC simulation in Fig.~\ref{PD}.
We comment that ``classical" MC simulations, which calculate the ensemble (time) average of physical quantities, cannot reveal such a dynamical oscillation of the electric flux explicitly.

On the other hand, in Fig.~\ref{sigma_t}(b) in the deep Higgs phase with $\tilde{J}=10$, the oscillation of $\sigma(t)$ is gradually lost, and $\sigma(t)$ converges to a constant value for any $\ell$.
This indicates the decay of the original flux structure along the line L$_{\pm}$, because a non-vanishing constant value of $\sigma(t)$ means that a fraction of the electric flux along L$_\pm$ disperses from the initial position and/or reduces in its strength from $E_0$. 
Due to the large $J\rho_0$ term, the energy escaped from L$_{\pm}$ diffuses 
into the entire space, i.e., ``condensation" of the gauge potential energy takes place.

\begin{figure}[t]
\centering
\includegraphics[width=8cm]{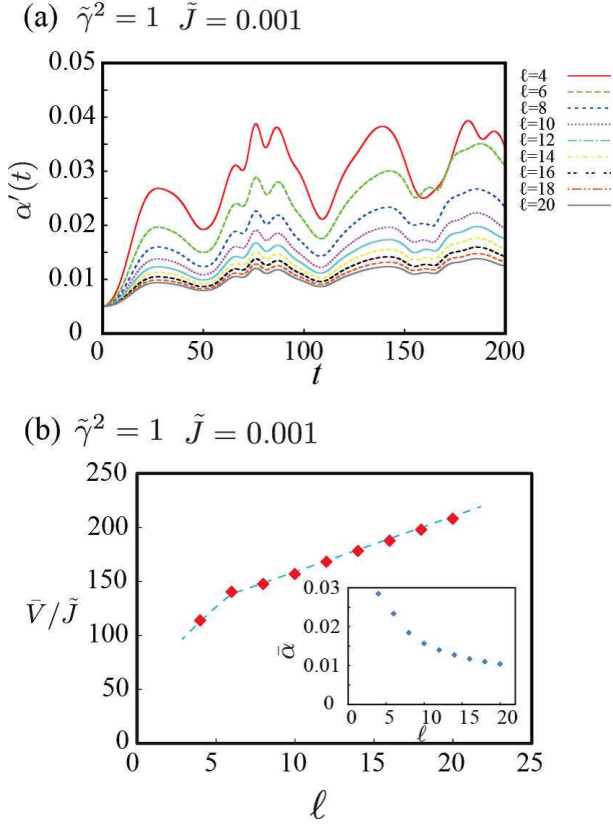}
\caption{(Color online) (a) Time evolution of the string tension $\alpha'$ of Eq.~(\ref{Vr}) for various lengths $\ell$ of the electric flux. 
(b) Time average $\bar{V}(\ell)$ of the potential $V(\ell)$ and time average $\bar{\alpha}$ of $\alpha'$ with the time interval from $t=0$ to $t=200$.
It is obvious that $\bar{\alpha}$ exhibits a smooth behavior converging to 
a constant as $\ell$ increases. 
}
\label{tension1}
\end{figure}


\begin{figure*}[t]
\centering
\includegraphics[width=16cm]{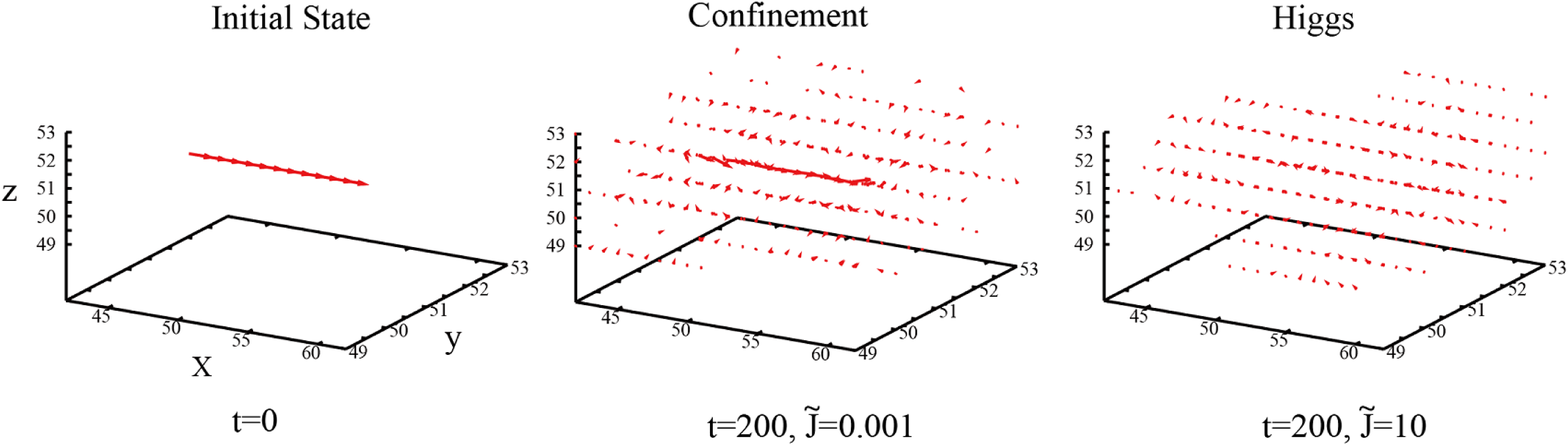}\\
(a) Snapshot of the electric field $E_{r,i}$ (red arrows)\\
\vspace{0.5cm}
\includegraphics[width=16cm]{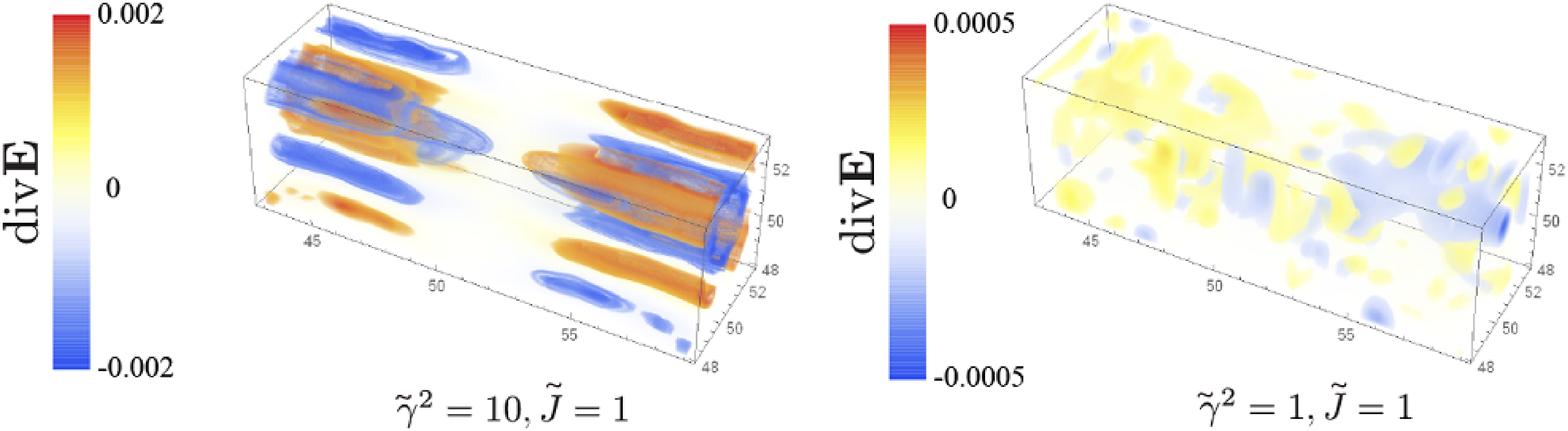}\\
(b) Snapshots of ${\rm div}{\bf E}_{r}\ (t=200)$
\caption{(Color online)
(a) Snapshots of the electric field $E_{r,i}$ for $\tilde{\gamma}^2=1$
with static charged sources pinned at $r_{\pm}$ (See the text) with the flux length $\ell=12$; (left) $t=0$, (center) $t=200$ for $z\tilde{J}=0.001$(confinement regime), (right) $t=200$ for $\tilde{J}=10$ (Higgs regime). 
The electric flux spanned between static sources clearly survives for $\tilde{J}=0.001$ and breaks for $\tilde{J}=10$. 
(b) Snapshots of the divergence of electric field ${\rm div}{\bf E}_{r}$ at $t=200$ with the same static sources as (a).
${\rm div}{\bf E}_{r}$ just measures the Higgs charge [See Eqs.~(\ref{dive}) and (\ref{dive2})], and its magnitude is certainly larger for $\tilde{\gamma}^{2}=10$ (left) than for $\tilde{\gamma}^{2}=1$ (right).}
\label{3DEfield}
\end{figure*}


In the confinement phase, it is interesting to measure the confinement
potential between the external charges $q_+$ and $q_-$.
In the confinement phase of LGTs without couplings to matter fields
the potential energy $V(\ell)$ of a pair of external charges (in the fundamental representation) separated by a distance $\ell$ is well fitted by a linear-rising confining potential, i.e., $V(\ell)=\alpha' \ell$, where $\alpha'$ is called the string tension \cite{wilson,ks}.
One may naively think that coupling to a matter field 
may change the potential to a short-range one because of the shielding effect.
However, the above observation of $\sigma(t)$ suggests that the confinement potential survives even in the presence of the Higgs field due to the energy exchange sketched in Eq.~(\ref{globaloscillation}). 
In fact, $\sigma(t)$ in Fig.~\ref{sigma_t} (a) shows that
the shape of the electric flux at $t=0$ is almost reproduced in every single period of the oscillation.

To study $V(\ell)$ in the confinement phase further, we measure the total energy $W(\ell)$ of the whole system with pinned down $q_+$ and $q_-$ and define $V(\ell)$ and the string tension $\alpha'$ as
\begin{eqnarray}
&& V(\ell)=W(\ell)-W(0) \;\;\; (\ell \gtrsim 2), \nonumber \\
&& \alpha'={V(\ell) \over \ell}.
\label{Vr}
\end{eqnarray}
Here we explain the time dependence of $W(\ell)$. 
After the update from $t$ to $t+\Delta t$, two values of the electric field 
$E_{r_+,1}$ and $E_{r_--1,1}$ attached to $q_+$ and $q_-$ change from their original value $E_0$ at $t$ to new values $E'_+$ and $E'_-$ according to Eqs.~(\ref{GPE}). 
This is an energy-conserving process.
Then we reset them as $ E_{r_+,1}(t+\Delta t)=E_{r_--1,1}(t+\Delta t) =E_0$ by hand for the next update.
This procedure certainly injects (or absorbs) the energy $\Delta W =
(V_0'/2)[2E_0^2-(E_+')^2-(E_-')^2]$ to (from) the system.
 
In Fig.~\ref{tension1} (a), we show the time evolution of $\alpha'$ for various $\ell$'s. 
As explained above, the total energy $W(\ell)$, and hence $\alpha'$, is not a constant of motion, and gradually increases with the oscillating behavior of Eq.~(\ref{globaloscillation}).
In Fig.~\ref{tension1} (b), we show the time averages $\bar{\alpha}$ and $\bar{V}(\ell)$ of  $\alpha'$ and $V$, respectively.
It is obvious that, as $\ell$ increases, $\bar{\alpha}$ monotonically decreases and tends to a constant. 
As a result, the potential exhibits an expected linear behavior $V(\ell) \propto \ell$, which strongly supports that the present system is in the confinement phase \cite{higgschargedyn}.

Note that the short-distance behavior for $\ell < 5$ in Fig.~\ref{tension1}
deviates from the linear dependence. 
This may come from the perturbative one-photon exchange effect that gives rise to the Coulomb potential like $V(\ell) \propto - {1/\ell}$ \cite{Rothebook}, although vacuum polarization by the Higgs field renormalizes the external charges. 
For small $\ell$'s, 
the confinement effect does not emerge significantly. 
This behavior can be seen in the data $\ell = 4$ of Fig.~\ref{sigma_t}, in which we cannot discriminate the behavior of $\sigma$ between (a) and (b). Two phases are distinguished by the long-range (large-$\ell$) behavior of $V(\ell)$.





Finally, in Fig.~\ref{3DEfield} we show the 3D electric field $E_{r,i}$ and the divergence ${\rm div}{\bf E}_{r}\equiv \sum_i\nabla_i E_{r,i}$.
The upper panels in Fig.~\ref{3DEfield} represent the snapshots of electric field $E_{r,i}$ at $t=200$ for the typical parameter values, $J=0.001$ (confinement) and $J=10$ (Higgs), where we measured the electric field in the vicinity of the initial electric flux. 
These results represent the intuitive picture expected from the LGT; 
the electric flux between static charges survives in the confinement phase, 
while it breaks off in the Higgs phase. 
In the lower panels of Fig.~\ref{3DEfield}, we show snapshots of the divergence of the 3D electric field $E_{r,i}$ for $\tilde{\gamma}^2=10$ and $1$ with $\tilde{J}=1$.
The value of the density plot of ${\rm div}{\bf E}_{r}$ for $\tilde{\gamma}^{2}=1$ is overall smaller than that for $\tilde{\gamma}^{2}=10$, that is, the Higgs charge in the weak Gauss-law coupling $\tilde{\gamma}^{2}=10$ is denser than that in the case $\tilde{\gamma}^{2}=1$. 
For $\tilde{\gamma}^{2}=10$, the distribution of ${\rm div}{\bf E}_{r}$ 
appears as a characteristic quasi-periodic structure. 
From these results in Fig.~\ref{3DEfield} (b), the expression ${\rm div}{\bf E}_{r}$ for Higgs charge depends significantly on the parameter $\tilde{\gamma}^{2}$. 
This result is qualitatively in good agreement with the previous expressions of Eqs.~(\ref{dive}) and (\ref{dive2}) in Sec II.

\section{Proposal for feasible experiment of cold atomic gases} \label{expproposal}

\begin{figure*}[t]
\centering
\includegraphics[width=16cm]{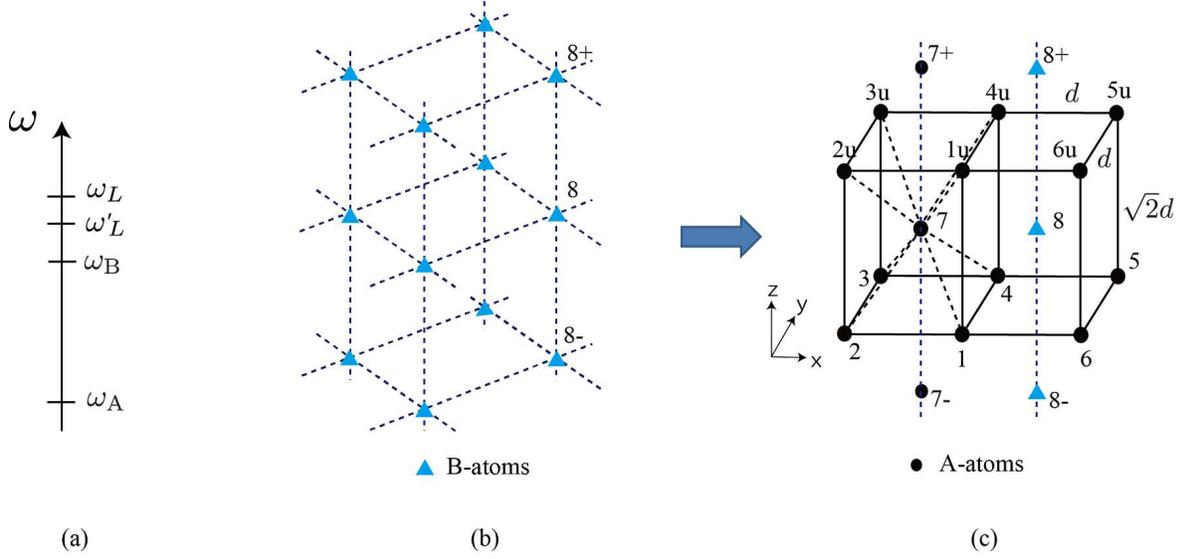}
\caption{(Color online) Setups to prepare the BCT optical lattice of Fig.~\ref{bct_lattice} with the interactions of Table I. 
(a) Choices of various frequencies $\omega$'s appearing in the trapping potentials of Eqs.~(\ref{vbct}) and (\ref{vocc}). 
(b) The cubic optical lattice built by the potential $V_{\rm OCC}$ of Eq.~(\ref{vocc}). Its sites (marked by triangles) are odd-column centers (OCC) such as the sites 8, 8$_{\pm}$ in Fig.~\ref{bct_lattice},
and occupied by B-atoms. (c) The BCT optical lattice built after switching the trapping potential from $V_{\rm OCC}$ to $V_{\rm BCT}$ of Eq.~(\ref{vbct}).  
The A-atoms reside on the sites marked by black circles,
while the B-atoms continue to occupy the OCC of (b).
}
\label{BCT2}
\end{figure*}

In this section, we propose a feasible experimental setup for realization of a cold atom system on a BCT optical lattice of Fig.~\ref{bct_lattice}, which is described by the extended Bose-Hubbard model of Eq.~(\ref{HEBH}). 
Then, as explained in Sec.~\ref{formulation}, under certain conditions such as uniform and large average atomic density, this atomic system is used to quantum-simulate the 3D gauge-Higgs model on a cubic gauge lattice.
The theoretical investigation given in Secs.~\ref{phasedia} and \ref{gpdynamics} may be a guide for such experimental simulations. 

It is possible to create a variety of lattice structure in 2D and 3D by appropriately arranging the propagation directions and the polarization of the laser beams \cite{Petsas,optical latticerev}.
To prepare a BCT optical lattice of Fig.~\ref{bct_lattice}, we follow the recent proposal by Boretz and Reichlof \cite{BCT}, and make use of the following optical potential
\begin{eqnarray}
V_{\rm BCT}&=&u\big[\cos^2(k_x x)+\cos^2(k_y y)+\cos^2(k_z z) \nonumber \\
&&+\cos(k_x x)\cos(k_y y)+\cos(k_y y)\cos(k_z z)  \nonumber \\
&&+\cos(k_z z)\cos(k_x x)\big],
\label{vbct}
\end{eqnarray}
where $k_{\alpha}=\pi/d_{\alpha}$ ($d_x=d_y=d$, $d_z=\sqrt{2}d$) and $x=n_x d_x$, $y=n_y d_y$, and $z=n_z d_z$ with $n_{\alpha} \in {\bf Z}$ ($\alpha = x,y,z$).
The form of Eq.~(\ref{vbct}) can be produced by the standard method with three pairs of counter-propagating laser beams.
We shall explain the coefficient $u$ later [See Eq.~(\ref{uab})].
One may check that the minima of $V_{\rm BCT}$ for $u < 0$ are located at sites of the BCT lattice. 
More precisely, there are two groups of minima; (i) $n_{\alpha} \in 2m_{\alpha}+1$ ($m_{\alpha} \in {\bf Z}$) corresponding to the center sites of the unit cells of the BCT lattice (such as the sites 7, 8 in Fig.~\ref{bct_lattice}) and (ii) $n_{\alpha} \in 2m_{\alpha} $ to corner sites of unit cells (such as 1$\sim$4, 1u$\sim$4u in Fig.~\ref{bct_lattice}).

To tune the atomic interactions in Table I, we propose the following two procedures; 

\nin
(D1) We prohibit cold atoms from occupying certain specific sites of the BCT optical lattice. 
Explicitly, we exclude atoms from the center sites of unit cells belonging to the odd-column [$(-)^{x+y}=-1$] such as 8, $8+$, and $8-$ in Fig.~\ref{bct_lattice}.
In what follows, we call these excluded sites odd-column centers (OCC).
This certainly satisfies the condition for group (iii) in Table~\ref{paratable}.

\nin
(D2) We adjust the parameters $J_{ab}$ and $V_{ab}$ so that the conditions for the groups (i), (ii), (iv), and (v) are satisfied. 
In fact, Rydberg atoms trapped on the optical lattice \cite{Rydbergatoms} 
may have an isotropic interaction with a $1/r^3$-type potential 
under a certain external electric field \cite{BuchlerHeidemann}. 
This long-range interaction is expected to satisfy 
the condition in the groups (i, ii).
 
The condition for the NNN interactions in the groups (iv, v) may be 
satisfied (without relying upon dipole-dipole interactions) by making use of extended anisotropic orbitals of Wannier states in the excited bands of an optical lattice.
In fact, this method was investigated explicitly in Ref.~\cite{ours2} to simulate the 2D gauge-Higgs model in a successful manner. We think that 
these methods are applicable also for the present 3D gauge-Higgs model without essential problems, and leave the details to be reported in a future publication.

Hereafter we focus our attention on the procedure (D1) above.
A scenario for (D1) to exclude the atoms in question, which we hereafter call {\it A-atoms}, from OCC is by introducing another kind of atoms, which we call {\it B-atoms}, and let them reside only on the OCC and give them strong repulsion to repel the original A-atoms.
The inter-species interaction Hamiltonian $\hat{H}_{\mathrm{AB}}$ between A-atoms and B-atoms is given by
\begin{eqnarray}
\hat{H}_{\mathrm{AB}} = U_{\mathrm{AB}}\sum_{c\in {\rm OCC}}{\hat \rho}_{c}{\hat n}_{\mathrm{B}c},
\end{eqnarray}
where  $c$ runs over the OCC, and ${\hat n}_{\mathrm{B}c}$ is the number operator of B-atoms residing on the site $c$.
The B-bosons are assumed to be in a Mott state, so ${\hat n}_{\mathrm{B}c}$ may be approximated by a uniform mean value, ${\hat n}_{\mathrm{B}c}\rightarrow \bar{n}_{\mathrm{B}}$. 
Then we have
\begin{eqnarray}
\hat{H}_{\mathrm{AB}} \simeq U_{\mathrm{AB}}\bar{n}_\mathrm{B}
\sum_{c\in {\rm OCC}}{\hat \rho}_{c}.
\end{eqnarray}
For sufficiently large $U_{\mathrm{AB}}\bar{n}_\mathrm{B}$, the probability 
that A-bosons reside on the OCC is suppressed significantly. 

From these consideration, we propose the following two steps to achieve the above procedure (D1);\\

\nin
{\it 1 Preparation of B-atoms on OCC.}

\nin
Start with a continuum harmonic trapping system including both A- and B-atoms.
Prepare a 3D optical lattice, the sites of which are just the OCC (See Fig.~\ref{BCT2} (b)) and occupied by B-atoms. It is a simple cubic lattice with the lattice spacing $\sqrt{2}d$. 
The corresponding optical lattice potentials $V'_\mathrm{A}$ and $V'_\mathrm{B}$ felt by A- and B-atoms, respectively, are given by
\be
V'_{\mathrm{A(B)}}&=&u'_{\mathrm{A(B)}}\left[\cos^{2}(k'_{1}x)+\cos^{2}(k'_{2}y)+\cos^{2}(k'_{3}z')\right],\nn
u'_{\mathrm{A(B)})} &=& - \frac{(d_{\mathrm{A(B)}}E')^2}{\hbar \Delta'_{\mathrm{A(B)}}} < 0,\nn
k'_i&=&\frac{\pi}{\sqrt{2}d},\ \Delta'_{\mathrm{A(B)}}=\omega'_{L}-\omega_{\mathrm{A(B)}},
\label{vocc}
\ee
where $E'$ is the electric field strength induced by a standing laser, $\omega'_L$ the laser frequency, and $\omega_{\mathrm{A(B)}}$ the ns-np energy gap of the A(B)-boson \cite{Zoller1}.
The amplitude $u'_{\mathrm{A(B)}}$ is the result of the second-order perturbation of the electromagnetic interaction between the instantaneous dipole $\vec{d}_{\mathrm{A(B)}}=q_{\mathrm{A(B)}}\vec{r}_{\mathrm{A(B)}}$ of A(B)-atoms and photons,
$(d_{\mathrm{A(B)}} E')^2\equiv\la (\vec{d}_{\mathrm{A(B)}}\vec{E}')^2\ra$ \cite{udashab}.  
$\Delta'_{\mathrm{A(B)}}$ is a detuning parameter for the A(B)-atom fixed on blue detuning, $\Delta'_{\mathrm{A(B)}} > 0$.
We choose $\omega'_L$ in such a way that $\Delta'_\mathrm{B} \ll \Delta'_\mathrm{A}$  (See Fig.~\ref{BCT2} (a)), so that the B-atoms are strongly trapped in OCC. \\

\nin  {\it 2 Changeover of optical potential.}

\nin
Switch the laser potential from $V'_{\mathrm{A,B}}$ of Eq.~(\ref{vocc}) to 
$V_{\rm BCT}$ of Eq.~(\ref{vbct}) within millisecond order.
Because of the mixture of A- and B-atoms, the amplitude $u$ in $V_{\rm BCT}$ becomes $u_\mathrm{A}$ and $u_\mathrm{B}$ for A- and B-atoms, respectively, which are given by
\be
u &\to& u_{\mathrm{A(B)}} \equiv - \frac{(d_{\mathrm{A(B)}}E)^2}{\hbar \Delta_{\mathrm{A(B)}}},\quad
\Delta_{\mathrm{A(B)}}=\omega_{L}-\omega_{\mathrm{A(B)}},
\label{uab}
\ee
where $E$ and $\omega_L$ are parameters of the standing laser after the switch.
As its time scale is smaller than the typical time scale of quantum tunneling between neighboring wells (Fig.~\ref{BCT2}(b)), this potential changeover may prevent the B-atoms from escaping from OCC.
Furthermore, by choosing $\omega_{L}$ so that 
\begin{eqnarray}
\Delta_{\mathrm{B}}\ll \Delta_{\mathrm{A}} \to |u_\mathrm{B}| \gg |u_\mathrm{A}|,
\label{detuning_ab}
\end{eqnarray}
 (See Fig.~\ref{BCT2} (a)), the B-atoms continue to stay on OCC, 
even though the A-atoms are allowed to tunnel into nearest neighbor sites. 

The resultant lattice system of A-atoms is shown in Fig.~\ref{BCT2}(c), which is described by $H_{\rm EBH}$ of Eq.~(\ref{HEBH}).
With the interaction parameters chosen according to the procedure (D2) above (and assuming $\bar{\rho}_0 \gg 1$),
this A-atom system is just described by $H'_{\rm EBH}$ of Eq.~(\ref{HEBH2})
or equivalently by $H_{\rm GH}$ of Eq.~(\ref{HGHM}).


\section{Conclusion} \label{concle}

In this section, we summarize the results of the paper and present our outlook.
In Sec.~\ref{formulation}, we started from the extended Bose-Hubbard model in the 3D optical lattice of Eq.~(\ref{HEBH}). 
Then we derived a low-energy effective model, Eq.~(\ref{HEBH2}), by assuming (i) a homogeneous and large average density $\la \hat{\rho}_a\ra =\rho_0 \gg1$ and (ii) small density fluctuations $\hat{\eta}_a (=\hat{\rho}_a-\rho_0)$ [the $O(\hat{\eta}^3)$-terms in Hamiltonian were neglected]. 
We showed that this effective model becomes equivalent to the gauge-fixed
version of the gauge-Higgs model in LGT when the interaction parameters $J_{ab}$ and $V_{ab}$ are suitably chosen, as in Table \ref{paratable}. This equivalence requires no special limit such as $\gamma \to 0$ owing to the inclusion of the Higgs matter field. 

We restricted ourselves to the region of $J$ and $\gamma^2$ that supports a 
uniform (site-independent) average density, $\la \hat{\rho}_a\ra =\rho_0$.
For other regions of parameters, the lowest-energy configuration may
favor an inhomogeneous pattern of $\la \hat{\rho}_a \ra$ supporting
density waves (See Appendix A). 
In fact, in a separate paper \cite{ours3}, we consider the extended Bose-Hubbard model in a one dimensional optical lattice for general values of the on-sight repulsion $V_0$ and the NN repulsion $V \equiv V_{ab}$. 
Among other things, we studied the phase diagram in the $V_0$-$V$ plane, and confirmed that it certainly includes the density-wave phase in which 
$\la \hat{\rho}_a\ra$ takes two alternative values on every other site.
Even in such a case, the equivalence to the gauge-Higgs model is maintained 
in some region of $V_0$ and $V$ by choosing $J_{ab}$ and $V_{ab}$ in a suitable manner.

In Sec.~\ref{phasedia}, we studied the phase structure of the gauge-Higgs model by a MC simulation.
The explicit phase diagrams of Figs.~\ref{PD} and \ref{PD2} may work as a guide for how to choose the model parameters in actual experiments of the system described by the extended Bose-Hubbard model.  
The coupling constants of the gauge-Higgs model are asymmetric in space-time directions, in contrast to the LGT models studied in high-energy physics. 
This point may open the possibility of a richer phase structure. 
The first-order phase transition we found is certainly such an example. 

In Sec.~\ref{gpdynamics}, we studied the time development of the extended Bose-Hubbard model by using the semiclassical Gross-Pitaevskii type approach.
Although the GPE underestimates the effect of quantum fluctuations and correlations, the obtained dynamical behavior of the electric field clearly changes as the coupling parameters change, reflecting the characteristics of each phase.
The location of a ``phase boundary" determined in this way is qualitatively consistent with the result of the static MC simulation of Sec.~III.
These approximate, but explicit and quantitative, solutions of the dynamical equation certainly help us, not only to design the actual set up of the experiments of quantum simulation, but also to gain a precise understanding of the real dynamics of the gauge theory. 
For example, to understand the structure of the potential of Fig.~\ref{tension1}, a simple shielding mechanism by pair creation of Higgs particles is not sufficient because the potential may saturate to a constant value when just a static Higgs pair is produced. 
The linear rising behavior at a larger distance may be understood by taking the kinetic energy of the Higgs bosons into account.
This is done by GPE, which respects the energy conservation law.

In Sec.~\ref{expproposal}, toward quantum simulation of the gauge-Higgs model, 
we present an explicit proposal to prepare an atomic system described by the extended Bose-Hubbard model. 
To realize the 3D gauge lattice, a BCT optical lattice is a suitable configuration. 
To prevent the occupation on OCC, one can use another kind of atoms to protect this occupation by a strong atom-atom repulsion. 
Adjustment of the NN and NNN interactions as in Table~\ref{paratable} seems 
to be a hard task, but engineering the atomic state into their higher-orbital state or suitably arranging the dipolar atoms or molecules can result in a desirable inter-site interaction, which may lead to a realization of the parameter setting of Table~\ref{paratable}. 

Our original aim is of course to simulate the target model, the gauge-Higgs model of LGT, by the base model, the extended Bose-Hubbard model of ultra cold atoms on the optical lattice.
However, the extended Bose-Hubbard model is an interesting model in its own regard, from a theoretical viewpoint and our understandings of it is far from complete. 
Our static and dynamical study of the gauge-Higgs model carried out in this paper
will certainly be of help to understand further the starting extended Bose-Hubbard model {\it at sufficiently low temperatures} because the gauge-Higgs model is derived as its low-energy effective model.
For example, the part $V_0' > 0$ of Fig.~\ref{PD}(c) is taken to describe the phase structure of the extended Bose-Hubbard model at large fillings and low temperatures.

Generally speaking, various notions and concepts established in LGT
find their places in understanding ultra-cold atomic systems on an optical lattice, and vice versa.
An explicit example of this mutual aid is discussed in Ref.~\cite{ours3},
where the Haldane-insulator phase in the 1D extended Bose-Hubbard model is interpreted by LGT.  
By stretching one's imagination in the opposite direction, the extended Bose-Hubbard model may shed some light to generalize LGT {\it beyond} the region of parameters where the present equivalence to LGT holds.

\acknowledgments
Y. K. acknowledges the support of a Grant-in-Aid for JSPS
Fellows (No.~15J07370).
This work was partially supported by a Grant-in-Aid
for Scientific Research from Japan Society for the 
Promotion of Science under Grant Nos.~26400246, 26400371, and 26400412.


\appendix

\section{Mean field calculation of the equilibrium atomic density in the 3D gauge lattice} \label{mftdenrho0}
In this appendix, we formulate a simple mean field theory to calculate 
the mean value of the atomic density $\rho_a\equiv \la \hat{\rho}_a \ra$ 
in the 3D gauge lattice shown in Fig.~\ref{bct_lattice}, where the atoms are located on the links of the gauge lattice. 
In particular, we are interested in the competition between the homogeneous state, in which $\rho_a $ is a site-independent constant, and an inhomogeneous state, which has some periodic distribution and supports density-wave excitations.

To derive the energy $E_{\rm MFT}$ of the mean field theory from $H_{\rm EBH}$ of Eq.~(\ref{HEBH}), we first set  $\exp(i\hat{\theta}_a)=1$ ignoring 
the fluctuations of the phase $\hat{\theta}_a$, and replace the amplitude operator by its average as $\hat{\psi}_a \to \sqrt{\rho_a}$ and $\hat{\rho}_a \to \rho_a$.
Then we obtain $E_{\rm MFT}$, including the chemical-potential term as
\be
E_{\rm MFT}(\rho) &=&-\sum_{a \neq b}J_{ab}\sqrt{\rho_a \rho_b}
+{V_0\over 4}\sum_a \rho_a (\rho_a-1) \nonumber \\
&&+\sum_{a \neq b}{V_{ab} \over 2} \rho_a \rho_b-\mu\sum_a \rho_a.
\ee
We consider the parameter setting shown in Table.~\ref{paratable}, $E_{\rm MFT}$ rewritten as 
\be
E_{\rm MFT}(\rho) &=& - J \sum_{a \neq b \in \mathrm{(i,ii)} }  \sqrt{\rho_a \rho_b}
+{V_0\over 4}\sum_a \rho_a (\rho_a-1) \nonumber \\
&& + {V \over 2} \sum_{a \neq b \in \mathrm{(i,ii,iv)}} \rho_a \rho_b - \mu\sum_a \rho_a.
\ee
The chemical potential $\mu$ is chosen as a function of the mean 
density $\rho_0 = \sum_{a=1}^{N_s} \rho_a / N_s$ over the sites, so that the total number of atoms in the system with $N_s (\equiv \sum_a 1)$ sites 
is a given number $N_s \rho_0$. 
In the case of a homogeneous density, we can take $\rho_a = \rho_0$, which is site-independent. 
Using this $\rho_0$, we rescale the energy as 
\be
\tilde{E}_{\rm MFT}(\zeta_a) &=& \frac{E_{\rm MFT}}{J \rho_0 }  \nonumber \\
&=& - \sum_{a \neq b \in \mathrm{(i,ii)} }  \sqrt{\zeta_a \zeta_b}
+{V_0 \rho_0 \over 4 J}\sum_a \zeta_a^2  \nonumber \nonumber \\
&& + {V  \rho_0 \over 2 J} \sum_{a \neq b \in \mathrm{(i,ii,iv)}} \zeta_a \zeta_b - {\mu \over J} \sum_a \zeta_a.
\ee
Here, we have introduced the scaled density $\zeta_a = \rho_a / \rho_0$ ($\zeta_a=1$ for the homogeneous case) and omitted the small $\rho_{a}^{-1}$ contribution in the on-site energy term. 
We minimize $\tilde{E}_{\rm MFT}(\zeta)$ with respect to $\zeta_a$ numerically to obtain an approximate configuration of $\zeta_a$ for the ground state.

Let us first consider the simplest case 
of vanishing NN and NNN couplings, $V=0$ ($\gamma^2=\infty$).
 Then the lowest-energy state with given $\rho_0$ is the uniform state $\zeta_a= 1$. This is understood because the inhomogeneous density fluctuations$\zeta_a=1+\delta\rho_a$ cost an extra energy $\Delta E$ as 
\be
&&\hspace{-1cm}\frac{\rho_0 V_0}{4J} \zeta_a^2 -\frac{\mu}{J} \zeta_a=
\frac{\rho_0 V_0}{4J} \left[ \left(\zeta_a-R\right)^2 - R^2 \right], \nn
R&\equiv& \frac{2\mu}{\rho_0 V_0},\nn
\Delta E&=&\frac{\rho_0 V_0}{4 J}\sum_a  \left[( 1 +\delta\rho_a-R)^2-(1-R)^2\right]\nn
&=& \frac{\rho_0 V_0}{4J}\sum_a(\delta\rho_a)^2 >0,
\ee
where $\sum_a\delta\rho_a=0$ due to the total atomic-number conservation.

\begin{figure}[ht]
\includegraphics[width=0.45\textwidth]{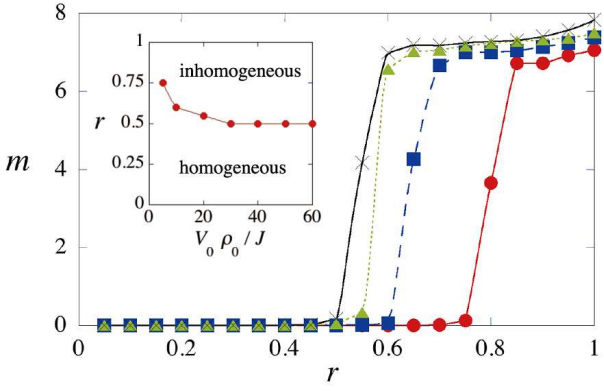}
\caption{Numerical result of the parameter $m$ of Eq.~(\ref{m}) as a function of the ratio of off-site and on-site repulsion,
$r\equiv \gamma^{-2}/V_0$ for $\rho_0 V_0 / J=5$ (circles), 10 (squares), 20 (triangles), and 30 (crosses). 
The parameter $m$ measures the non-uniformness of the ground state. 
The inset shows the parameter space revealing the boundary between 
the homogeneous and inhomogeneous density distributions. 
For sufficiently large $\rho_0 V_0 / J$, the density distribution of the ground state is homogeneous for $r \alt 0.5$ and becomes inhomogeneous for $0.5 \alt r$. }
\label{figm}
\end{figure}
For the opposite case of vanishing on-site coupling $V_0=0$, the lowest energy state can be determined so that it minimizes the inter-site coupling energy. 
Intuitively, this term dislikes the homogeneous density distribution, 
because, if we assume alternative density undulations such as $\rho_a = \rho_0 + \delta \rho$ and $\rho_b = \rho_0 - \delta \rho$, the simple inequality $(\rho_0+\delta\rho)(\rho_0-\delta\rho) =\rho_0^2-(\delta\rho)^2 < \rho_0^2$ implies that the system prefers the density wave state. 


From these considerations, we expect that the lowest energy state is
homogeneous when the ratio of $V ( = \gamma^{-2})$ and $V_0$, 
$r \equiv V/V_0 = \gamma^{-2}/V_0$ is sufficiently small, while it 
becomes inhomogeneous as $r$ becomes sufficiently large. 
To measure the degree of inhomogeneity of the lowest energy state, we use
\be
m \equiv \frac{1}{N_s} \sum_{a=1}^{N_s} \left( \zeta_a -1\right)^2,
\label{m}
\ee
which is zero for the homogeneous state and increases for the inhomogeneous one.
The value of $m$ can be calculated for general $r$ and $V_0 \rho_0/J$ by minimizing $E_{\rm MFT}$ numerically with respect to $\zeta_a$.
In Fig.~\ref{figm} we plot $m$ as a function of $r$ for several $V_0 \rho_0/J$.
It certainly supports our expectation above. 
We note that, as the value $\rho_0 V_0/J$ increases, the critical ratio $r$ saturates the value 0.5. 
From this result and the relation $V_0' = V_0 - 2\gamma^{-2}=V_0(1-2r)$,
positiveness of the electric energy ($V_0' > 0$) implies a homogeneous ground state ($r < 0.5$) and vice versa.
The inhomogeneous density distribution forms nontrivial patterns because of the intrinsic complexity of the 3D gauge lattice. This will be reported and discussed elsewhere.

\section{Details and supplementary results of the MC calculations}\label{mc}
In the MC calculations in Sec.~\ref{phasedia}, we use the standard Metropolis algorithm \cite{metropolis}, which typical sweeps 50000 (thermalization) + 10 (samples) $\times$ 5000 (measurement) in a single run for a fixed set of $c_i$'s, and calculate errors as the standard deviation of the 10 samples. 
We take the gauge-invariant expression  $\tilde{Z}_{\rm GH}$ of Eq.~(\ref{ZGHinv}) and {\it update both the gauge field $\theta_{x,\mu}$
and the Higgs field $\varphi_x$}. 
This process is known to accelerate the convergence of the Markov process more than using the gauge-fixed expression $Z_{\rm GH}$ of Eq.~(\ref{4DGHM}) would.
The linear sizes of lattice $L$ we used are $8,12,16,20,24$ and the phase transition points of Fig.~\ref{PD} are determined by the data of $L=16$. Typical acceptance ratios are $0.7\sim 0.8$. 

\begin{figure}[t]
\centering
\includegraphics[width=7cm]{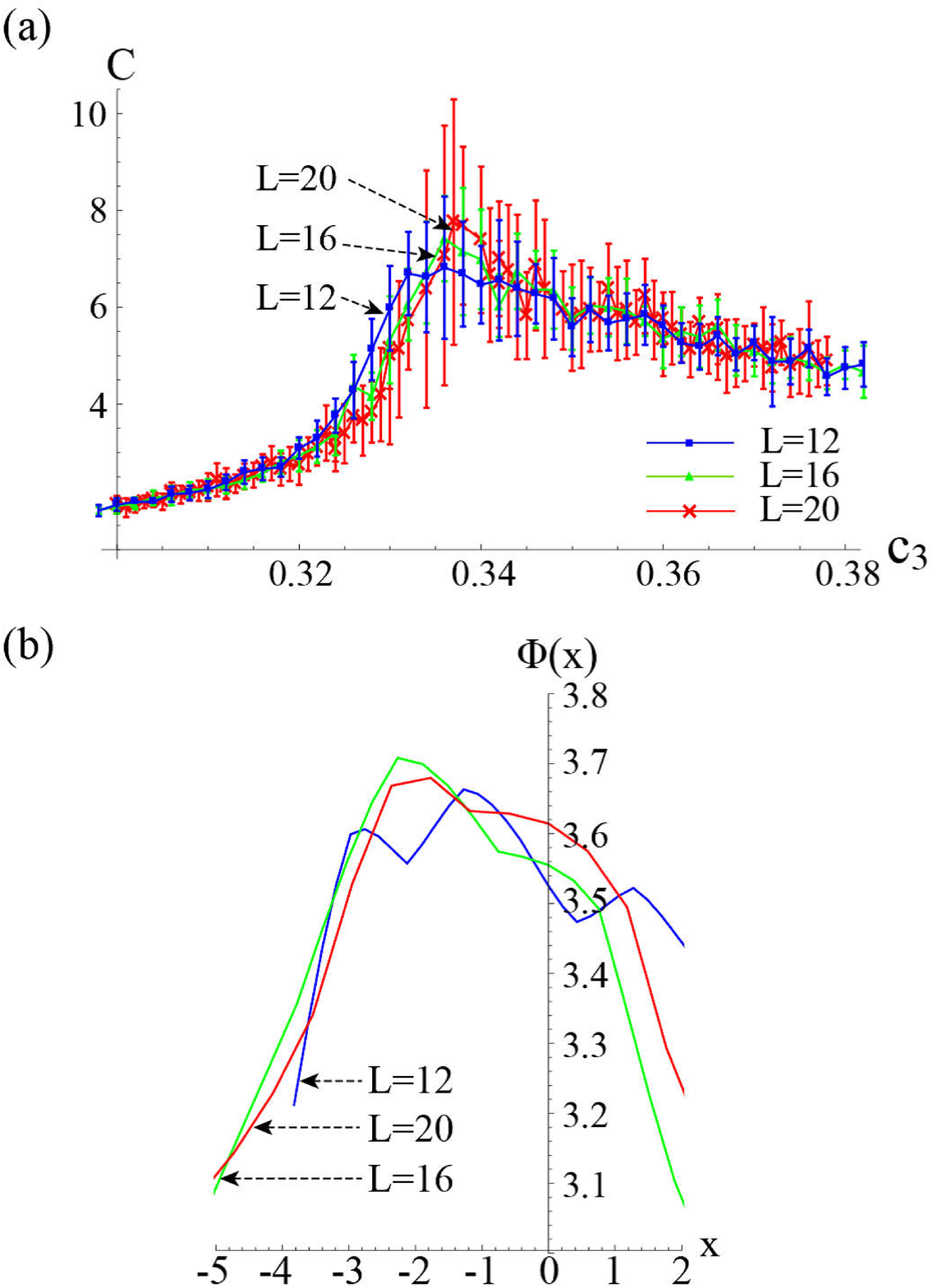}
\caption{(Color online) (a) Specific heat $C$ vs.\ $c_3$ for $L=12,16,20$
at $c_1=0.0$ and $c_2=0.4$.
(b) Scaling function $\phi(x)$ in Eq.~(\ref{scaling2}) determined from the data (a).
The scaling parameters in Eq.~(\ref{scaling2}) are determined as $\sigma=0.125,\ 
\nu=0.50,\ c_{3c}=0.339$.}
\label{scaling}
\end{figure}

For a second-order transition, one expects that the specific heat $C(T)$ obeys 
the following finite-size scaling behavior \cite{compphys}
 \be 
 C(T)=L^{\sigma/\nu}\phi(L^{1/\nu}\epsilon),\quad \epsilon=\dfrac{T-T_c}{T_c},
 \label{scaling2}
 \ee
 for sufficiently large $L$, where $T_c$ is the transition point at $L\to \infty$.
In Fig.~\ref{scaling}(a) we show the typical size-dependence of $C(c_3)$ at $c_1=0, c_2=0.4$ for $L=12,16,20$ around a second-order transition
point $c_3 \simeq 0.33\sim 0.34$.
As $L$ increases, the three curves $C(c_3)$ seem to vary systematically as
 expected by Eq.~(\ref{scaling}).
In Fig.~\ref{scaling}(b), we show the scaling function $\phi(x)$ determined by using the data of Fig.~\ref{scaling}(a). The optimal values of the critical exponents and critical $c_3$ are determined as $\sigma=0.125,\ \nu=0.50$, and $c_{3c}=0.339$, respectively.
Be taking the errors in $C(c_3)$ in Fig.~\ref{scaling}(a) into account, 
we think that the three curves in Fig.~\ref{scaling}(b) define an approximate scaling function $\phi(x)$.
This supports that the systems with $L \gtrsim 12$ are approximately in the scaling region.

To check whether the number of sweeps (5000 in our case) for the measurement of a sample is large enough, one may use the integrated autocorrelation time $\tau(N)$ \cite{compphys}, where $N$ is the total sweeps for measurement ($N=10\times 5000$ in our case) and the function $\tau(k)\ (1\leq k \leq N)$ is defined for an observable $O$ as
 \be 
\tau(k)&\equiv& \frac{1}{2}+\sum_{t=1}^k\frac{f(t)}{f(0)},\quad
\bar{O}\equiv \frac{1}{N}\sum_{i=1}^NO(i),
\nn
f(t) &\equiv& \frac{1}{N-t}\sum_{i=1}^{N-t}\left(O(i+t)-\bar{O}\right)\left(O(i)
-\bar{O}\right),
\label{tau}
\ee
where $O(i)\ (i=1,\cdots,N)$ is the value at the $i$-th sweep after the thermalization.

As $O$, we use the internal energy $U$ of Eq.~(\ref{UC}). As $k$ increases, 
$\tau(k)$ saturates to a constant and then oscillates slightly. 
The saturated values (defined by the first local maximum) 
for $c_1=0.0, c_2=0.4$ are shown in Table II.\\

\begin{table}[t]
\centering
\caption{\label{autocorrelation} Saturated values of the integrated autocorrelation time $\tau(k)$ of Eq.~(\ref{tau}) calculated by the data for the internal energy of a single run with $c_1=0.0$ and $c_2=0.4$.
}

\begin{tabular}{|c|c|c|c|c|c|c|c|}\hline
$c_3$&0.32&0.325&0.33&0.335&0.34&0.345&0.350\\ \hline
$L=12$&61& 86 &238 &235 &180 &281&115  \\ \hline
$L=16$&59&90  &410&441   &174 &119& 79 \\ \hline
$L=20$&68&65  &424& 922  &222&106  & 85\\ \hline
\end{tabular}
\end{table}

\nin
As expected, the saturated values become larger as $C(c_3)$ approaches its peak.
However, all the values are well below the number of sweeps for measurement of each sample, 5000. 
We judge that 5000 is large enough to truncate auto correlations and define independent samples.   

These results may support that the essential structure of the global phase diagrams Figs.~\ref{PD} and \ref{PD2} are reliable as the size dependence of the peak in Fig.~\ref{scaling}(a) is tiny compared with the scale of these diagrams.
 For a more precise determination of location of phase transition points,
 we certainly need more intensive calculations with high statistics and larger lattices.

\section{Derivation of expectation value of the electric field Eq.~(\ref{we})} \label{electricterm_Eq18}
In this appendix we derive Eq.~(\ref{we}). 
We start from the first line of Eq.~(\ref{zelecric}), which is written as 
\begin{eqnarray}
\exp\biggl[\biggr(-\sum_{x,i}\frac{E^{2}_{x,i}}{2c_2}+iE_{x,i}\theta_{x,0i}\biggr)\biggr].
\label{electricex1}
\end{eqnarray}
Then, by introducing a source $J_{x,i}$ for the electric field $E_{x,i}$, 
the generating functional of $Z_{\rm GH}(J)$ is obtained as
\begin{eqnarray}
\hspace{-0.7cm}
&&Z_{\rm GH}\rightarrow Z_{\rm GH}[J]= \int [D\theta_{x,\mu}] [DE_{x,i}]\nn
&&\exp\biggl[\sum_{x,i}\biggr(-\frac{E^{2}_{x,i}}{2c_2}+iE_{x,i}(\theta_{x,0i}+J_{x,i})\biggr)+ \cdots\biggr]\nn
&\propto& \int [D\theta_{x,\mu}] \exp\biggl[\sum_{x,i}c_{2}\cos(\theta_{x,0i}+J_{x,i})+
\cdots\biggr],
\label{electricex3}
\end{eqnarray}
where $\cdots$ denotes the terms $A_{\rm I}$ and $A_{\rm P}$ that are independent of $E_{x,i}$, and the last line is obtained 
by summation over $E_{x,i}$ following the step to Eq.~(\ref{zelecric}).
Then we obtain the expectation value $\la E_{x,i}^n \ra$
by letting the partial derivative $(-i\partial /\partial J_{x,i})^n$ act on $Z_{\rm GH}[J]$.
As a result we obtain the variance of the electric field of Eq.~(\ref{we}) as 
\begin{eqnarray}
\hspace{-0.7cm}
&&\langle(E_{x,i}-\langle E_{x,i}\rangle)^{2}\rangle =  \langle E_{x,i}^2\rangle-\langle E_{x,i}\rangle^2\nn
&&=\biggl[-i\frac{\partial }{\partial  J_{x,i}}Z_{\rm GH}[J]\biggr|_{J_{x,i}\rightarrow 0}\biggr]^{2}
-\biggl[-\frac{\partial ^2}{\partial  J_{x,i}^{2}}Z_{\rm GH}[J]\biggr|_{J_{x,i}\rightarrow 0}\biggr]\nn
&&=c_{2}\langle \cos\theta_{x,0i}\rangle -c^{2}_{2}\langle \sin^{2}\theta_{x,0i}\rangle.
\label{electricex4}
\end{eqnarray}

\section{Interpretation of the phase diagram} \label{interprephasedia}

In this appendix, let us discuss the global structure of Fig.~\ref{PD},
the phase diagram of the 3D gauge-Higgs model given by Eq.~(\ref{4DGHM}), and interpret the order of transitions by a plausible argument.

First, we consider the case $c_2=0$. Then, the total action $A_{\rm GH}$ of Eq.~(\ref{4DGHM}) decouples to $A_{\rm I}$ of the time-like $\theta_{x,0}$ and $A_{\rm L}$ of the space-like $\theta_{x,i}$, and $Z_{\rm GH}$ becomes
\be
Z_{\rm GH}|_{c_2=0}&=&(Z_{c_1})^{L^4}(Z_{\rm 3D XY})^L,\nn
Z_{c_1} &=&\int_{-\pi}^\pi \frac{d\theta}{2\pi}\exp(c_1 \cos\theta)=I_0(c_1),\nn
Z_{\rm 3DXY}&=&\int [D\theta_{r,i}]
\exp\big[c_3\sum_{r, i<j}\cos(\theta_{r,i}-\theta_{r,,j})+\cdots.\big].\nn
\ee
The integrals over $[D\theta_{x,0}]$ decouples to $L^4$ sites, and each site gives rise to $Z_{c_1}$, the modified Bessel function.
$Z_{c_1}$ has no singularity in $c_1$, and gives an average $\la U_{x,0}\ra=\la\cos\theta_{x,0} \ra =I_1(c_1)/I_0(c_1)$, which starts from 0 at $c_1=0$ and increases as $c_1$ increases up to 1.
The integrals over $[D\theta_{x,i}]$ decouple to $L$ spatial 3D gauge lattices labeled by $x_0$, and each 3D system gives rise to $Z_{\rm 3DXY}$, the partition function of the 3D XY spin model. 
This is because the $c_3$ term in $A_{\rm GH}$ at fixed $x_0$ is just the
energy of the NN XY spin model $E_{\rm 3DXY}=-c_3\sum_{(a,b) \in (i)}\vec{S}_a \cdot \vec{S}_b$ of spins $\vec{S}_a =(\cos\theta_a,\sin\theta_a)$, defined on a 3D optical lattice with the identification $\theta_{r,i}\leftrightarrow \theta_a$ as in Fig.~\ref{bct_lattice} of Sec.~\ref{formulation}.
$Z_{\rm 3D XY}$ is known to exhibit a second-order phase transition
at $c_3 = c_{3c}\simeq 0.34$. 
For $c_3 > c_{3c}$ there is an order of $\theta_{r,i}$ and disorder otherwise.
The horizontal second-order transition curve for $c_2=0$ in Fig.~\ref{PD} expresses just this transition where the critical value of $c_3$ has no $c_1$ dependence because $Z_{c_1}$ is analytic.     

Next, we consider the effect of the $c_2$ term, $A_{\rm P}$, which couples
$\theta_{x,0}$ and $\theta_{x,i}$. In the mean-field type interpretation,
one may decouple it as follows;
\be
&&c_2\ U^\dag_{x,0}U^\dag_{x+0,i}U_{x+i,0}U_{x,i}\to
c_2' U^\dag_{x+0,i}U_{x,i}
+c_2'' U^\dag_{x,0}U_{x+i,0},\nn
&&c_2'\equiv c_2\la\ U^\dag_{x,0}U_{x+i,0}\ra,\quad
c_2''\equiv c_2\la U^\dag_{x+0,i}U_{x,i}\ra.
\label{c2mft}
\ee
The first term in the R.H.S. of Eq.~(\ref{c2mft}) is the NN pair of the ``XY" spin  $U_{x,i}$ in the $\mu=0$ direction with a ``coupling constant" $c_2'$. 
So this term and $A_{\rm L}$ compose the ``pseudo"-4D XY model of XY spins $U_{x,i}$.  
Of course, this is not a genuine 4D XY model because its coupling $c_2'$
is ``soft"; it contains fluctuations of another variable $U_{x,0}$.  
For sufficiently large $c_1$, the $A_{\rm I}$ term prepares a saturated value $\la U_{x,0}\ra \sim 1$ with small fluctuations.
So $c_2'$ is almost a stable constant and the system becomes almost a genuine 4D XY model with asymmetric couplings $(c_3, c_2')$.
This model is known to exhibit a second-order phase transition as its 3D counterpart, irrespective of the value of $\tilde{c}_3$ as long as it is a constant.
This explains the second-order transitions at large $c_1$ in Fig.~\ref{PD}. 
The second term of Eq.~(\ref{c2mft}) is the NN coupling of {\it time-like} XY spin $U_{x,0}$ in the 3D lattice at fixed $x_0$ with a soft ``coupling constant", $c_2''$. 
This gives rise to a set of $L$ decoupled 3D XY spin models, each of which is labeled by $x_0$. The term $A_{\rm I}$ works
as an external source to $U_{x,0}$. 

Therefore, the total system with the replacement (\ref{c2mft}) is the sum of two subsystems; (i) one 4D XY model with coupling $(c_3,c_2')$ and 
(ii) $L$ 3D XY models with coupling $c_2''$ and the source.   
Through the soft couplings $c_2', c_2''$ these two subsystems affect each other.
For example, let us start with the phase where both $U_{x,i}$ and $U_{x,0}$
are disordered, i.e., small  $c_2', c_2''$.
If $c_2'$ develops once by fluctuation, $U_{x,i}$ spins favor ordering, 
which, in turn, may increase $c_2''$ and favor an ordering of $U_{x,0}$ and lead to larger $c_2'$. 
That is, $c_2'$ and $c_2''$ rapidly increase each other by a synergistic effect. 
This is in strong contrast with the usual ``hard" coupling constants. 
As one changes the usual constants $c_1, c_3$ with fixed $c_2$, $c_2'$ and $c_2''$ may not change linearly with $c_1, c_3$, but stay at zero until a certain critical point is reached and then rise continuously but abruptly. 
This behavior of soft couplings certainly brings the would-be second-order 
transition to a first-order transition. This is one explanation of the 
first-order transition shown in Fig.~\ref{PD}. 
The conditions to achieve the above scenario of first-order transitions are 
(i) sufficiently large $c_2$ and (ii) sufficiently small $c_1$,
because (i) $c_2'$ and $c_2''$ are proportional to $c_2$ and the above
synergistic effect needs a certain amount of sensitivity for each other,
and (ii) if $c_1$ is large enough, $\la U_{x,0}\ra$, and hence $c_2'$ has a small fluctuation and behaves almost as a ``hard" constant.
 
%


\begin{thebibliography}{99}
\bibitem{gamma}
In Refs.~\cite{Zohar1,Zohar2},  it is proposed that one tunes the 
repulsive intersite interaction strength 
for nearest-neighbor sites and a subset of next-nearest neighbors to a common value 
$\gamma^{-2}$  (See Table I and \changed{Eq.~(\ref{HEBH2})} in Sec.~\ref{formulation}) and takes the limit 
$\gamma^{-2} \to \infty\ (\gamma^2\to 0)$.
\bibitem{mc3+1d}
We used MC simulations of the corresponding LGT defined 
on the 3+1-dimensional (3+1)D lattice
where the extra one dimension is the imaginary-time axis in the path-integral
representation of the zero-temperature partition function \cite{wilson}.
\bibitem{nnvsnnn}
For the case that $V_{ab}$ of the group (iv) has a
different value from $\gamma^{-2}$ of the groups (i) and (ii),
by repeating similar calculations we still obtain a model of LGT,
but it contains long-range interactions between the vector potentials.
It may be an interesting gauge theory worth to be studied, but we shall 
confine ourselves below to the standard gauge model with 
local interactions.
\bibitem{scaling0}
The finite-size scaling analysis in Appendix B shows that
the $L$-dependence of the critical values such as $c_3$ at a transition point
is much smaller than the difference between points on different 
phase-boundary curves with different $c_2$ in Fig.~\ref{PD} [See Fig.~\ref{scaling} (a)].
Therefore, the data of Fig.~\ref{PD} calculated in $L=16$ (supported to be in the scaling 
region in Appendix B) are sufficient for the following discussion. 
The more precise data with larger $L$ as well as the data with different values of $c_2$
are welcome to refine the study of convergence (extrapolation) as $c_2\to\infty$.
\bibitem{negativevprime}
The electric energy term with $V_0' < 0$ induces 
 unstable fluctuations of $E_{x,i}$, which may drive the atomic system to inhomogeneous 
 density configurations such as the density-wave state \cite{ours3} (See Appendix 
 \ref{mftdenrho0}).
 Although one may define a gauge-Higgs model even in such a case,
 it may have quite different properties from those of familiar LGT models.
\bibitem{higgschargedyn}
Although at a first glance one may expect the formation of a Yukawa-like potential between two static charges 
in the Higgs phase, such a potential was not measured in the simulations 
because the strong fluctuations of the electric field surrounding the initial electric flux 
drowns out such an exponentially decaying potential.
\end{thebibliography}
\end{document}